\pdfoutput=1
\RequirePackage{color}
\RequirePackage{ifpdf}

\documentclass{JINST}

\usepackage{color}
\usepackage{url}

\title{The Photomultiplier Tube Calibration System of the MicroBooNE Experiment}

\author{J. Conrad$^a$, B.J.P. Jones$^a$, Z. Moss$^a$, T. Strauss$^b$\thanks{Corresponding author.} ~and M. Toups$^a$\\
\llap{$^a$}Massachusetts Institute of Technology, 77 Massachusetts Avenue, Cambridge, MA 02139, USA
\llap{$^b$}University of Bern, LHEP, Sidlerstrasse 5, CH-3012 Bern, Switzerland\\
E-mail: \email{thomas.strauss@lhep.unibe.ch}}

\abstract{We report on the design and construction of a LED-based fiber calibration system for large liquid argon time projection detectors. This system was developed to calibrate the optical systems of the MicroBooNE experiment.  As well as detailing the materials and installation procedure, we provide technical drawings and specifications so that the system may be easily replicated in future LArTPC detectors.}

\DeclareGraphicsExtensions{.pdf,.png,.jpg}


\usepackage{lineno}

\begin{document}

\section{Introduction}
\begin{figure}[h]
\centering
\minipage{0.55\textwidth}
\includegraphics[width=\linewidth]{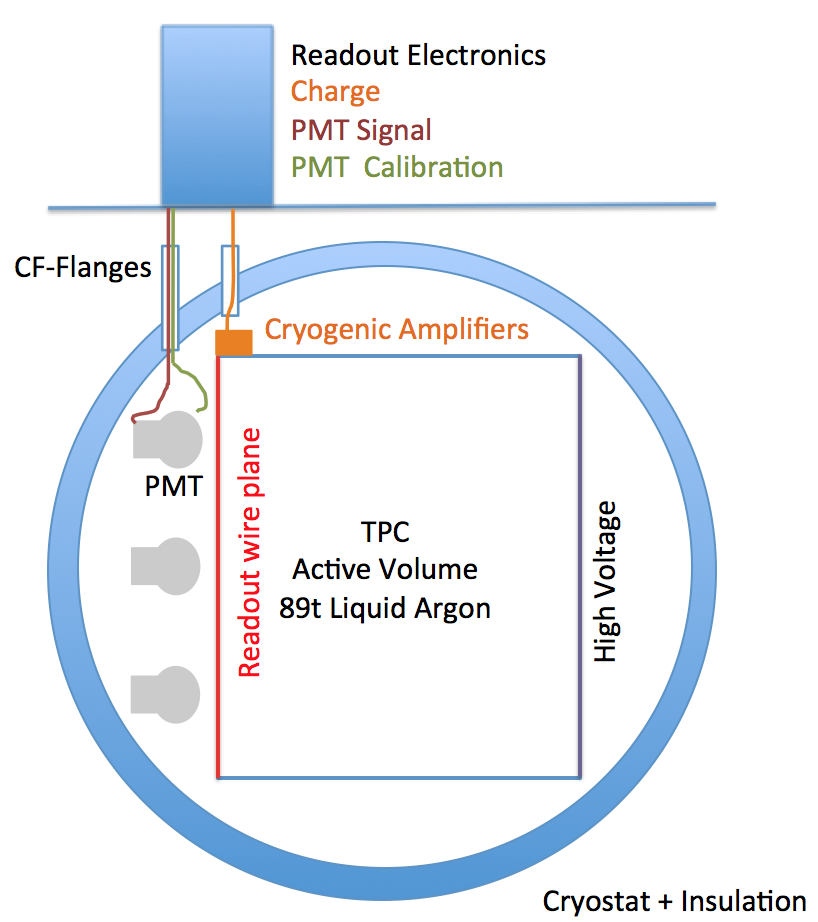}
\endminipage\hfill
\caption{Layout of the MicroBooNE detector. Inside the cryostat is the liquid argon TPC which has an active volume of 89\,t. The charge is read-out by wires, the scintillation light is recorded with PMT's. Feed-throughs connect the cryogenic detector parts to the readout electronics outside.\label{fig:detector layout}}
\end{figure}

The MicroBooNE experiment at Fermilab is investigating the nature of the low energy neutrino excess observed in the MiniBooNE experiment at the Booster Neutrino Beam (BNB) line. At a similar distance as MiniBooNE, it will be able to identify the electromagnetic nature of the excess by providing an excellent electron-photon identification. The experiment is using the liquid argon time projection chamber (LArTPC) to observe neutrino interactions. As shown in Figure \ref{fig:detector layout} the detector uses a cryostat to hold 170\,tons of liquid argon, of which an active volume of 89 ton is readout by application of negative 128\,kV on the TPC cathode, leading to a drift of electrons from ionizing particles crossing the active volume towards readout wire planes. Here the signals from the drifting electrons will be amplified and transferred to readout electronics outside the cryostat, where intermediate amplifiers and ADC allow further processing and event reconstruction. The ionizing particles will also produce scintillation light in the UV, which can be readout out with 32 $\times$ 8-inch Hamamatsu R5912-02MOD photomultiplier tubes (PMTs) \cite{Briese:2013wua}. Argon scintillates in the far ultraviolet, so tetraphenyl butadiene (TPB) coated plates are used to shift the light into a visible wavelength. Each PMT is housed inside a magnetic shield to minimise the effect of gain losses due to the earths magnetic field on the position of the first PMT dynode. The timing signal provided by the PMT readout will allow to determine which reconstructed tracks is associated to the neutrino event by synchronisation with the BNB beam window, and which tracks are induced by cosmic rays. For R\&D purposes, four 2-inch Hamamatsu R7725-MOD PMTs connected to TPB-coated light guides \cite{Baptista:2012bf} are also installed.  

In this report we document the design and installation of this calibration system. In Section \ref{sec:design} we summarise the design considerations. Section \ref{sec:optics} describes the optical and mechanical components and Section \ref{sec:electronics} outlines the design and testing of the electronics for the system. Finally we provide technical drawings and specifications in Appendix \ref{app:A} and \ref{app:B} so that variants of the system can be easily replicated for future cryogenic TPC detectors.

\section{Design Considerations \label{sec:design}}
\begin{figure}[t]
\centering
\minipage{0.70\textwidth}
\includegraphics[width=\linewidth]{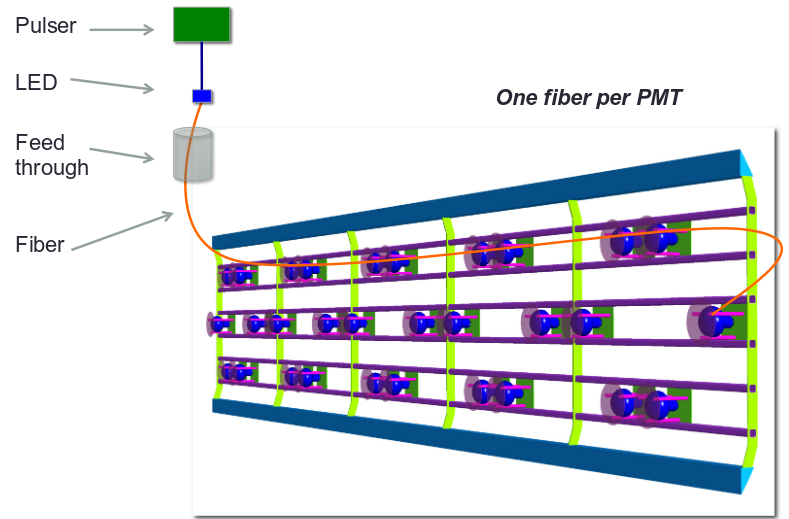}
\endminipage\hfill
\caption{Layout of the PMT calibration system detector. Optical fibers connect each PMT inside the MicroBooNE cryostat to a feedthrough. Outside the cryostat, the fibers extend to an electronic LED driver board to allow the single photon pulse calibration.  \label{fig:calibration layout}.}
\end{figure}

\begin{figure}[t]
\centering
\minipage{0.55\textwidth}
\includegraphics[width=\linewidth]{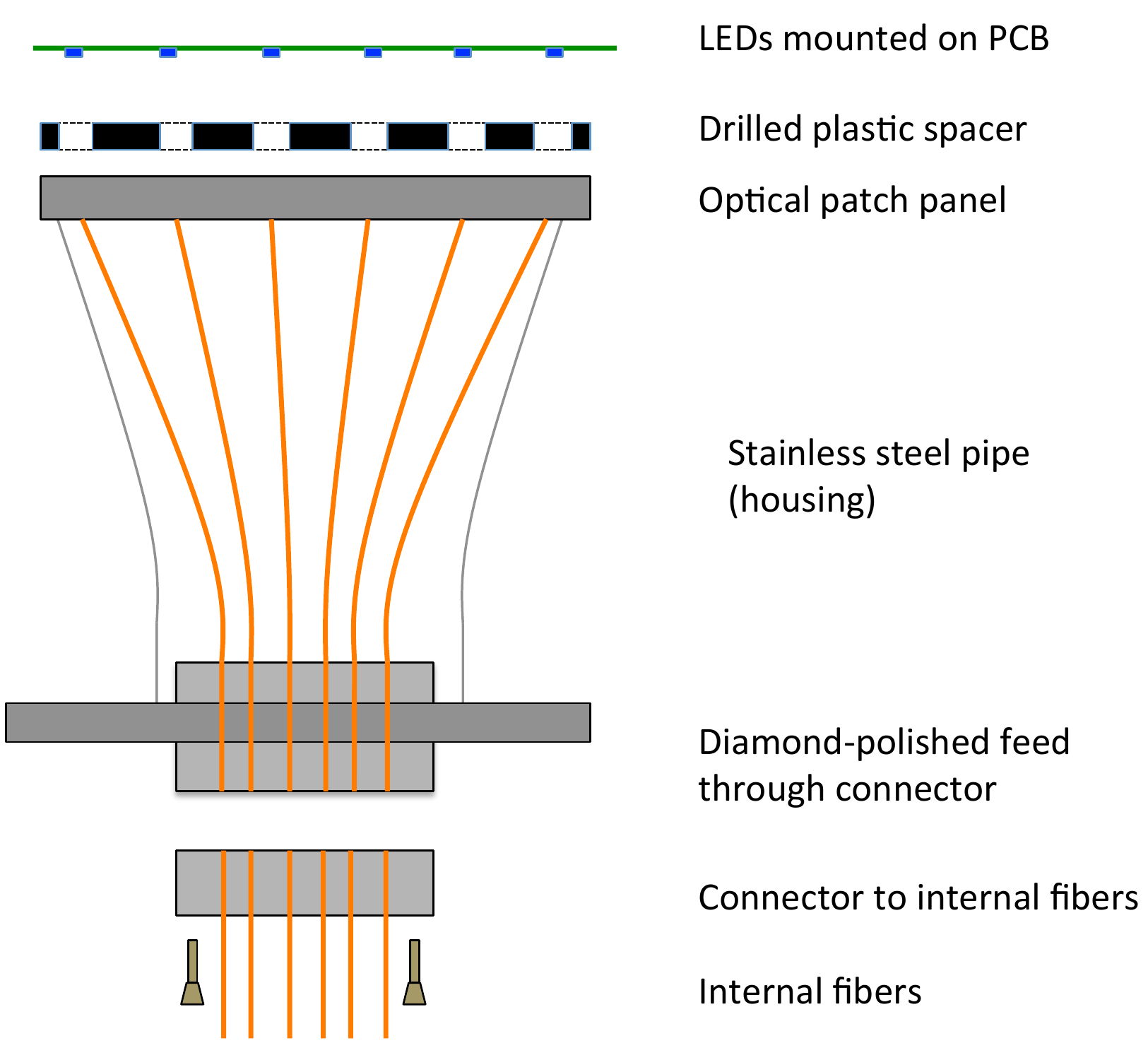}
\endminipage\hfill
\minipage{0.40\textwidth}
\includegraphics[width=\linewidth]{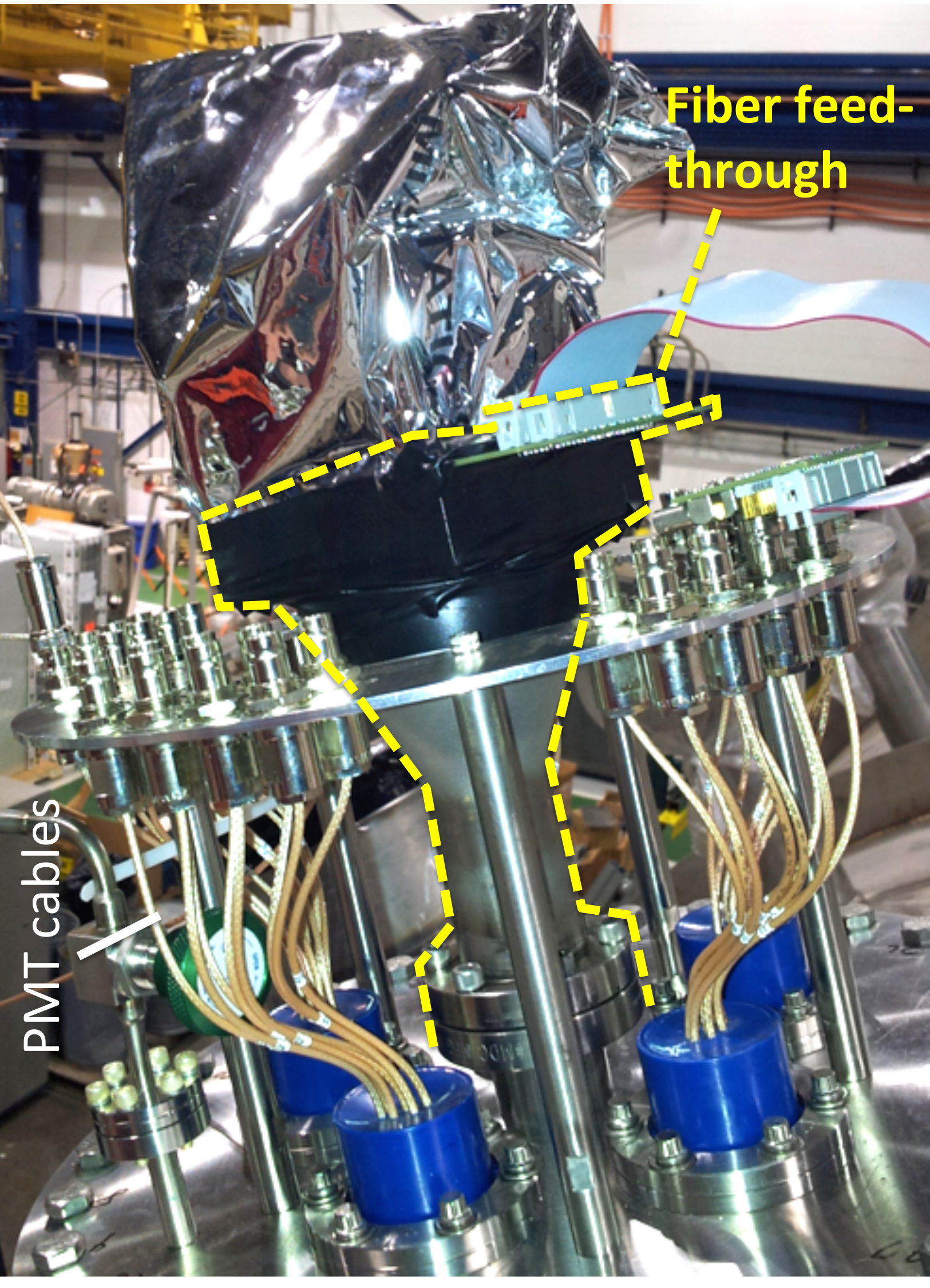}
\endminipage\hfill
\caption{Left: Sketch of the custom feed-through design. The internal fiber connector couples to the feed-through from below. Outside the flange, the fibers are spread onto a patch panel which couples to the LED array. Right : Photograph of the feed-through installed on the cryostat, as used during the PMT installation tests \label{fig:ben_cartoon_edit}.}
\end{figure}

The goal of the PMT calibration system is to check the timing of the PMT system, exercise the optical system during construction and commissioning, and to calibrate it once the detector is operational. For this purpose a LED-driven fiber calibration system was developed. Shown in Figure \ref{fig:calibration layout} is the schematic of the system, a control board pulses an array of LEDs, each of which is coupled to an optical fiber.  Each PMT has an individual calibration fiber, which runs from the LED array outside the detector, through a custom cryogenic feed-through (pictured in Figure \ref{fig:ben_cartoon_edit}), to the PMT.  

The setup divides into two components, the cryostat internal and outside components; 'inside' are 10\,m long optical fibers and their mounting connectors, 'outside' is the fiber feed-through and the control boards. This is due to the partial installation of the system inside the MicroBooNE cryostat, whose layout is shown in Figure \ref{fig:detector layout}. The cold components are optical fibres and a mounting connector to direct the LED light pulse onto the PMT surface. Similar fibers to those deployed in this system were repeatedly cryo-cycled in test stands and do not show any significant fatigue.  Operation in a dedicated purity test stand showed no observable effects on liquid argon free-electron lifetime \cite{pordes}. Due to the installation procedure they had to be installed in advance of the TPC installation inside the cryostat, and before extensive work outside the cryostat took place, involving a 3.5\,km move of the cryostat on the Fermilab side, application of insulating foam and the installation of a platform to support the electronic readout crates. 

Outside the cryostat is the feed-through connector and control board, which need to be accessible for future maintenance and possible exchange on failure. Due to limited space on the available cryostat feed-throughs, all optical fibers had to be fitted within a custom-made flange, mounted on the MicroBooNE PMT system feed-through. Space constraints require that the fibers penetrate a ConFlat-type (CF) flange with a 2.75-inch inner diameter \cite{cf}. This flange had to fulfill the pressure vessel requirement of the cryostat with 3\,bar. Additionally the feed-through needed to be vacuum tight for the potential cryostat evacuation and must work at cryogenic temperatures.

All LEDs on the control board can be pulsed either individually or simultaneously, for gain and timing calibrations, respectively.  The system is designed to have a timing precision of a few hundred picoseconds (the recorded PMT timing resolution), and studies using prototypes have demonstrated gain stability measurements of better than 1\% (\cite{Jones:2013bca} for single photon LED pulses. The LEDs are flashed in response to a logic pulse supplied by the MicroBooNE data-acquisition system with a frequency of a few kHz.  This allows the optical system timing and gain to be re-measured between periods of physics data taking.  Both timing and gain stability are important properties of the optical system, which functions as a trigger,  and as the primary cosmogenic background rejection tool for MicroBooNE.

\section{Optical and mechanical components \label{sec:optics}}
A single calibration fiber extends from the 400\,nm LED on the LED driver board (mounted outside the cryostat on the feed-through) to approximately 2 cm in front of the photocathode at the PMT inside the detector. The fibers penetrate the cryostat through a custom-made flange, as described in Section \ref{sec:design}.

Figure \ref{fig:ben_cartoon_edit} shows a sketch of the custom feed-through design. Internal fibers that each run to one PMT are bundled and set into a connector, which is coupled using screws onto the inside of the feed-through. A second set of fibers bundled into a matching connector on the inside of the flange are spread onto a patch panel which couples to the LED array. The LED array is affixed directly onto the top of the feed-through and operates in warm conditions outside the cryostat. The break in the fibers at the feed-through was implemented (a) to avoid damaging the delicate fibers during installation, and (b) to allow for exchange of the flange if required.  We successfully tested the vacuum tightness under cryogenic conditions as well as at 3\,bar overpressure found no leakage. Technical drawings of the LED flasher feed-through can be found in the Appendix \ref{app:A}.

\begin{figure}[b]
\centering
\minipage{0.33\textwidth}
\includegraphics[width=\linewidth,height=3cm]{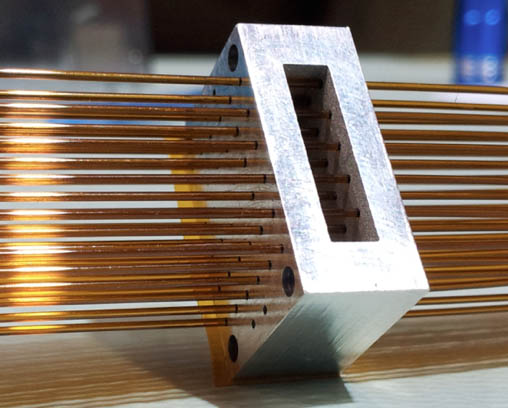}
\endminipage\hfill
\minipage{0.33\textwidth}
\includegraphics[width=\linewidth,height=3cm]{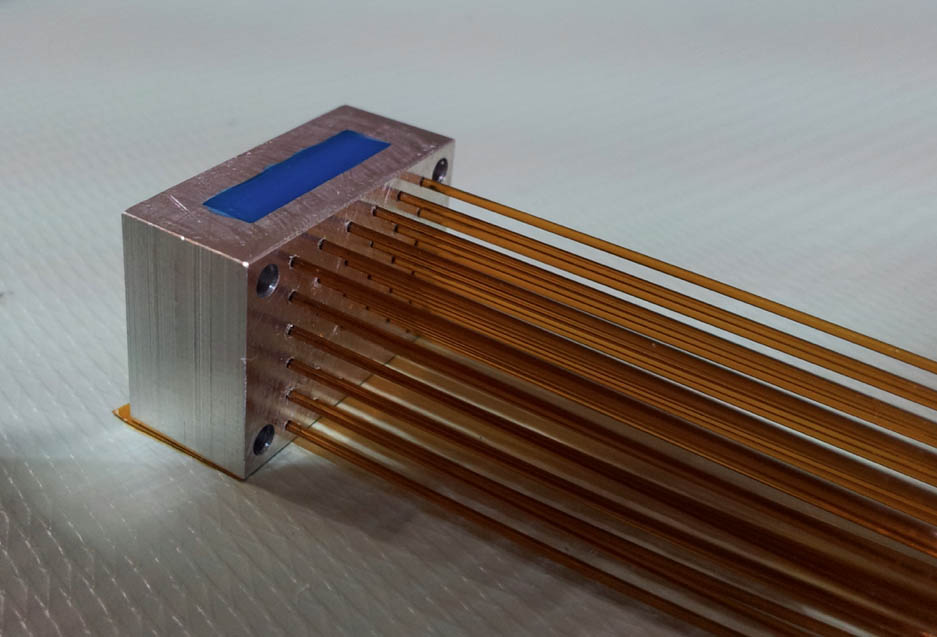}
\endminipage\hfill
\minipage{0.33\textwidth}
\includegraphics[width=\linewidth,height=3cm]{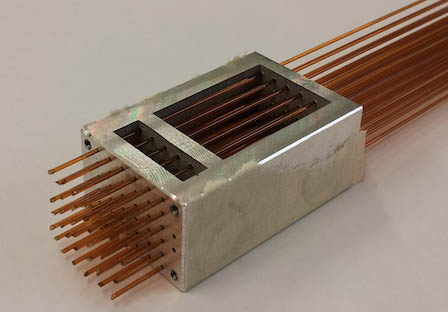}
\endminipage\hfill
\caption{Preparation of the two connector parts.  Left and center : internal connector.  Fibers are drawn through, cut using a fiber cleaver, aligned, set in sealant and than diamond polished. Right :  Feed-through connector with fibers drawn through. \label{fig:connectors}.}
\end{figure}

\begin{figure}[t]
\centering
\minipage{0.35\textwidth}
\includegraphics[width=\linewidth]{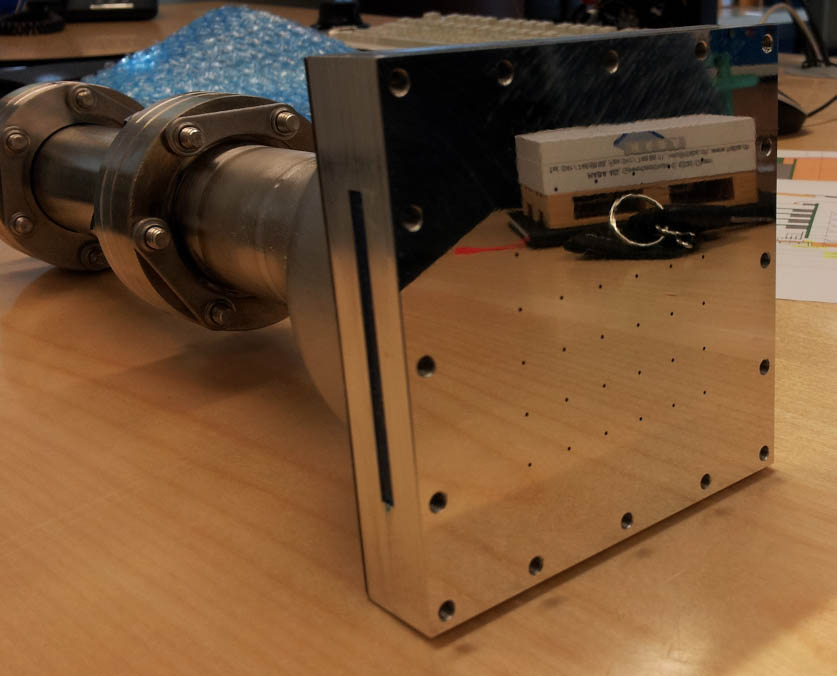}
\endminipage\hfill
\minipage{0.63\textwidth}
\includegraphics[width=\linewidth]{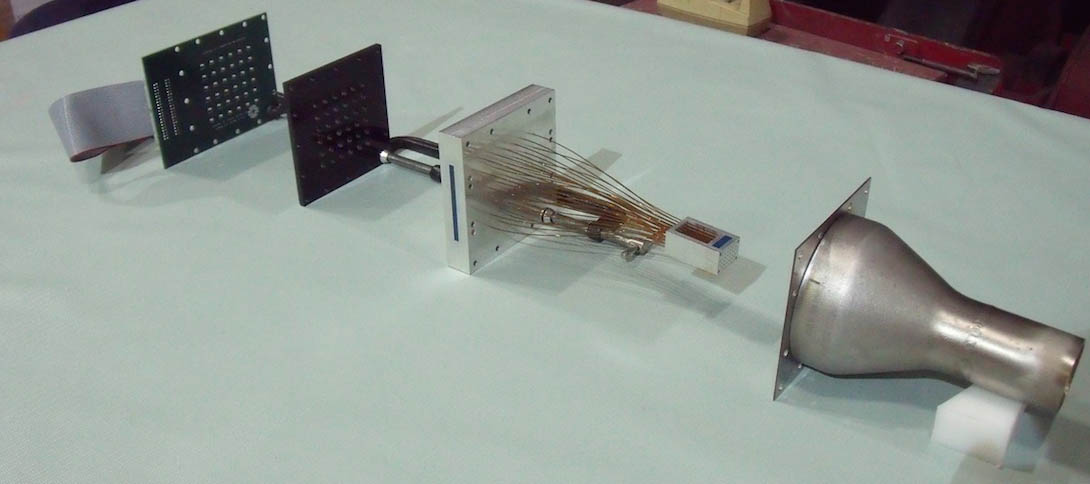}
\endminipage\hfill
\caption{Left : the assembled feed-through during the vacuum test, after diamond polish.  Right : custom feed-through components, exploded view (compare with the schematics Figure 3). The fibers are inserted into the optical patch-panel and feed-through connector block, on the right the steel cone of the feed-through is visible, on the left the drilled plastic spacer and the LED array board \label{fig:screen09}.}
\end{figure}

\begin{figure}[h]
\centering
\minipage{0.23\textwidth}
\includegraphics[width=\linewidth,height=5.25cm]{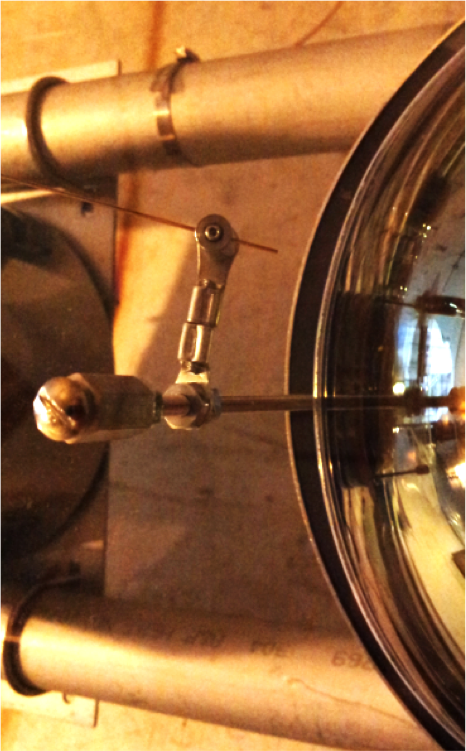}
\endminipage\hfill
\minipage{0.19\textwidth}
\includegraphics[width=\linewidth,height=5.25cm]{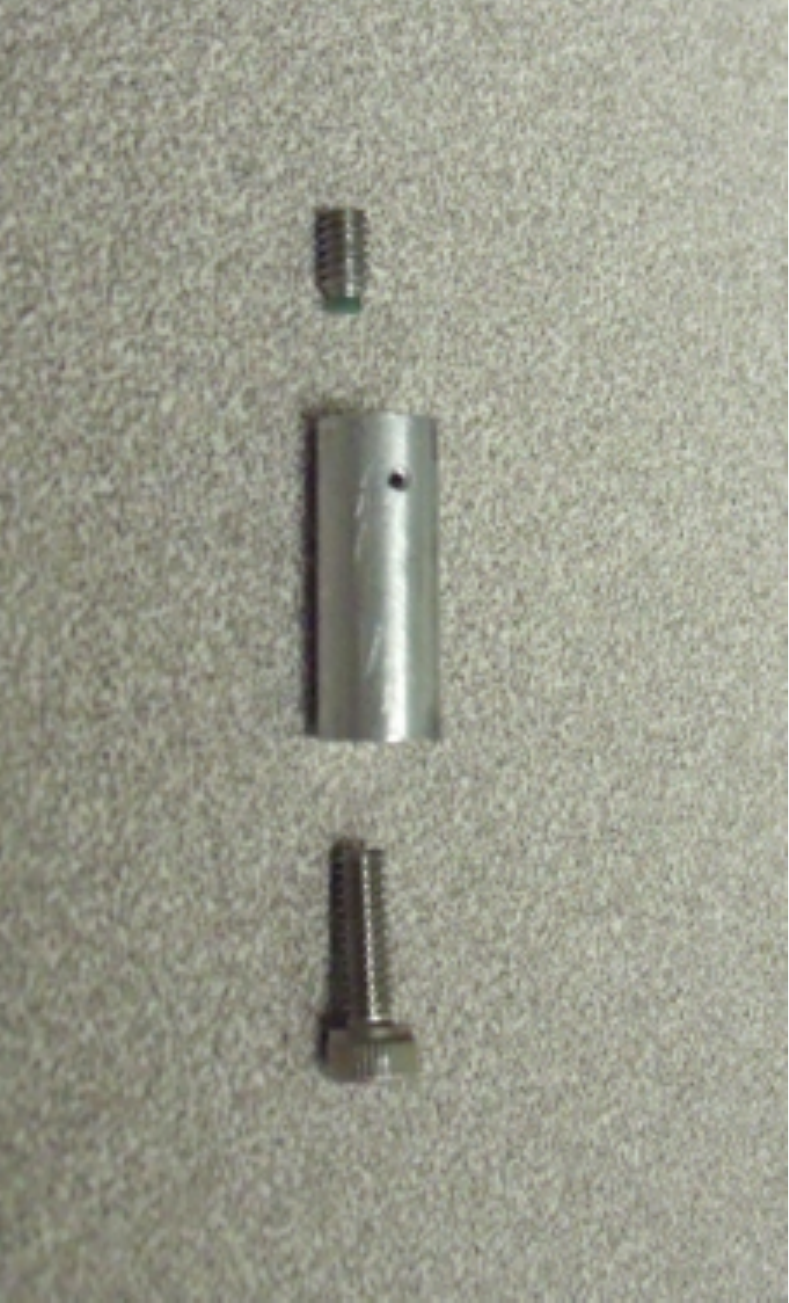}
\endminipage\hfill
\minipage{0.28\textwidth}
\includegraphics[width=\linewidth,height=5.25cm]{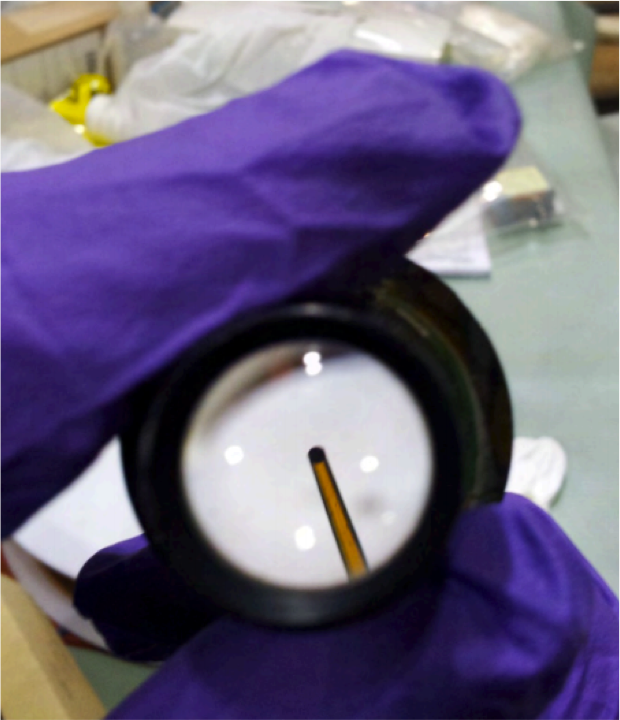}
\endminipage\hfill
\minipage{0.25\textwidth}
\includegraphics[width=\linewidth,height=5.25cm]{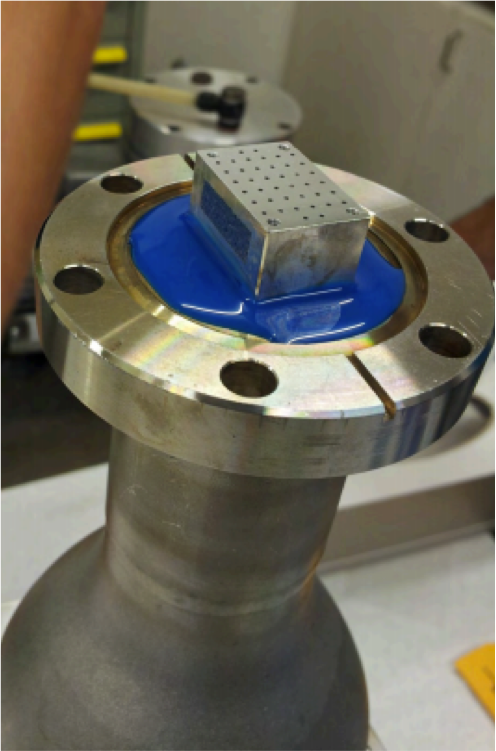}
\endminipage\hfill

\caption{Left to right : the mounting stand-off which holds the fiber at each PMT ; the aluminium standoff with its nylon-tipped set-screw ; magnification of the prepared fiber end, which points towards the photocathode ; connector block being set in the custom feed-through. \label{fig:MountParts}}
\end{figure}

In the coupling blocks the fibers are arranged in a $6\times6$ matrix with grid spacings of 1.5 mm.  Both the internal block and the feed-through block are shown in Figure \ref{fig:connectors}. For the internal block, after routing the long fibers to their respective PMTs, each fiber end was inserted into the connector. Their ends were prepared with a fiber cleaver, then aligned, and then the hole was filled with polymer sealant to hold the fibers in place.  For the feed-through, the short fibers were routed between the patch panel and connector block. They were then set using the small hole in the connector block (see Figure \ref{fig:connectors}, right)  without their ends being prepared.  The full assembly was placed into the CF flange penetration, and the larger hole was filled, sealing the flange  (Figure \ref{fig:MountParts}, right).  The polymer flow around the fibers is important to provide a high quality seal.  Both sides of the feed-through were then diamond polished (Figure \ref{fig:screen09}, left).  All fiber ends were thus prepared as high quality optical surfaces.

The sealant material used here is Arathane CW 5620 blue with HY 5610 hardener \cite{Arathane}. The malleable Arathane was chosen to secure the fiber ends without damaging the fibers through thermal expansion and contraction, following previous successful use in liquid argon test stands \cite{jones}.

Figure \ref{fig:screen09} right shows an exploded view of the CF feed-through assembly, the schematics can be found in Figure \ref{fig:ben_cartoon_edit}. Short fibers are already attached to the optical patch-panel and the feed-through connector block. Left from the optical patch panel are shown the drilled plastic spacer (black) and the LED array board (green, with attached ribbon cable), which is further described in Section \ref{sec:electronics}. The components fit into a stainless steel pipe which provides mechanical protection as well a light-tight environment.  The connector block is then set using Arathane into a CF flange, which is welded onto the steel pipe (not shown).  After assembly, the surfaces of the patch panel and the fiber connectors were diamond polished to optimize the quality of optical coupling between the fibers.

We use Molex FVP polymide fibers \cite{molex} with diameter of 600\,$\mu$m, cladding of 30\,$\mu$m and an additional buffer layer of 25\,$\mu$m. Each fiber connecting a PMT to the flange has a length of 13.84$\pm$.03\,m. The fiber attenuation at 400nm is less than 0.3\,dB/m (room temperature). The fibers are routed along the PMT frame inside the detector and held in place by teflon ties, with strain relief at few-meter intervals along the fiber path. Since they have no outer jackets, these fibers are very fragile, and so great care was taken during installation to avoid snags or kinks.  The fibers are flexible when in pristine condition, but they become weak and brittle if the buffer layer is scratched. To avoid damage to the buffer layer after installation, each contact point of the fibers along the frame was padded with teflon netting. 

The internal fibers are each mounted to individual PMT assemblies by a fiber holder: an aluminium standoff with a nylon-tipped set screw as shown in Figure \ref{fig:MountParts}. Each fiber holder is attached to a threaded rod which holds the TPB plate away from the magnetic shield. The fiber holders connect to these rods via two ring tongue terminals (Panduit clamps) crimped together onto a 1\,mm threaded rod. clamped in place by two nuts and a lock washer. The end of each fiber points at the PMT photocathode from underneath the wavelength shifting plate, approximately 2\,cm from the surface of the PMT.	

\section{Electronic components \label{sec:electronics}}
\subsection*{Description}
\begin{figure}[t]
\centering
\includegraphics[width=0.65\textwidth]{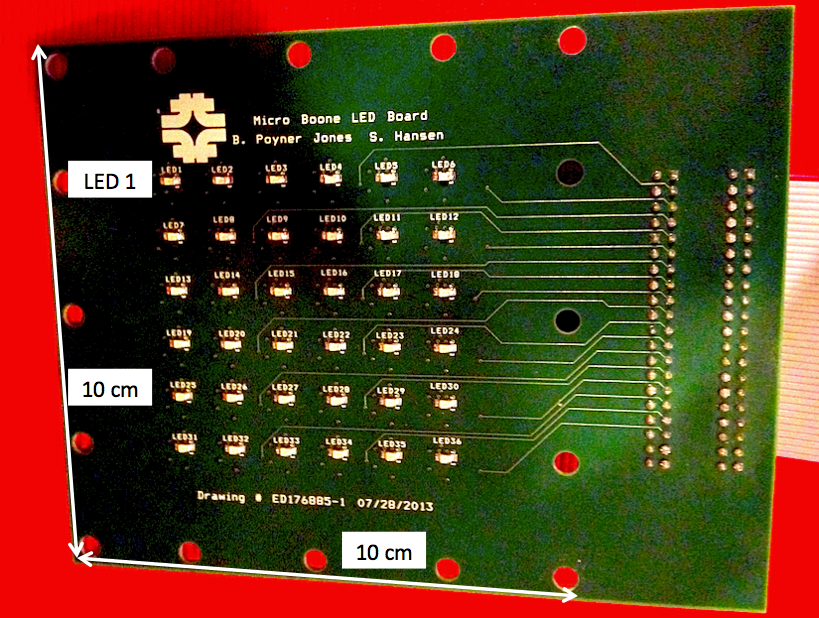}
\includegraphics[width=0.65\textwidth]{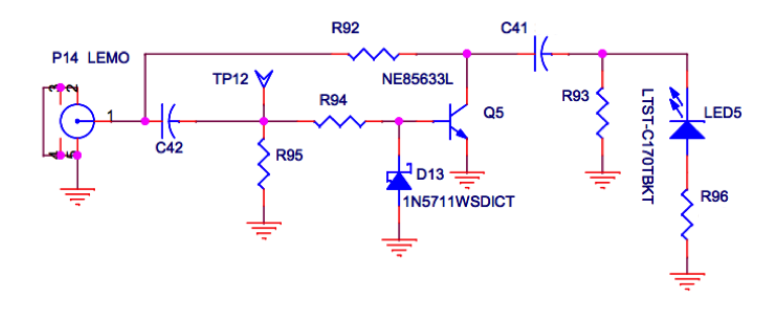}
\caption{Top: An image of the LED board with the 400\,nm LEDs mounted. Bottom : A single LED network schematic from LED flasher board. \label{fig:flashboard}}
\end{figure}

The driving electronics for the flasher LEDs \cite{Sten} are based on a system which has been used to pulse a single LED at a liquid argon test stand \cite{jones}. The flasher system electronics consist of two boards: the LED board and the driver board.
The LED board (Figure \ref{fig:flashboard}) is an array of 36 pulser networks placed to the back of the board, each of which is connected to a 400\,nm LED mounted on the front surface.  The LEDs are arranged in a 6$\times$6 grid with 1\,cm spacings. The board is mounted to the optical patch panel, where each LED aligns with a single fiber. To accommodate the LED thicknesses, and to isolate each LED-fiber interface, a plastic spacer with 3\,mm wide holes is sandwiched in between the board and the patch panel, as shown in Figure \ref{fig:ben_cartoon_edit}. The back of the board is covered by a drilled plastic housing screwed down onto the top side of the patch panel. The use of 16 screws provides a tight seal, and additional black silicon will be used to ensure light tightness. The PCB layout and schematic for the LED board can be found in the Appendix \ref{app:B} in Figures \ref{fig:LEDBOARDpcb} and \ref{fig:LEDBOARDsch} respectively.

A schematic diagram of one of the pulser networks is shown in Figure \ref{fig:flashboard}, bottom. The network accepts two inputs through the port (P14) : a transient trigger signal and a DC bias. When no transient is present at the input, (Q5) is not conducting, and the DC bias charges the simple RC circuit comprised of (R92), (C41), and (R93). When the circuit is triggered, the transient passes (C42) and biases (Q5) into conduction. When (Q5) switches as described, a low-impedance path appears between (C41) and ground. A short pulse of current will flow through (LED5) which is now forward biased, emitting a brief pulse of light.

Although more complicated than a direct coupling of transient to the LED, this circuit has the benefit of weak reliance on the amplitude and width of the transient input. As long as the pulses are large enough to bias Q5 into conduction, wide enough to drain C41 completely, and infrequent enough to allow C41 to charge in between triggers, the light output will be stable and the numerical results will be discussed in detail below. The output amplitude, then, depends only on the DC offset voltage, which can be controlled precisely with a DAC.

\begin{figure}[t]
\centering
\minipage{\textwidth}
\includegraphics[width=\linewidth]{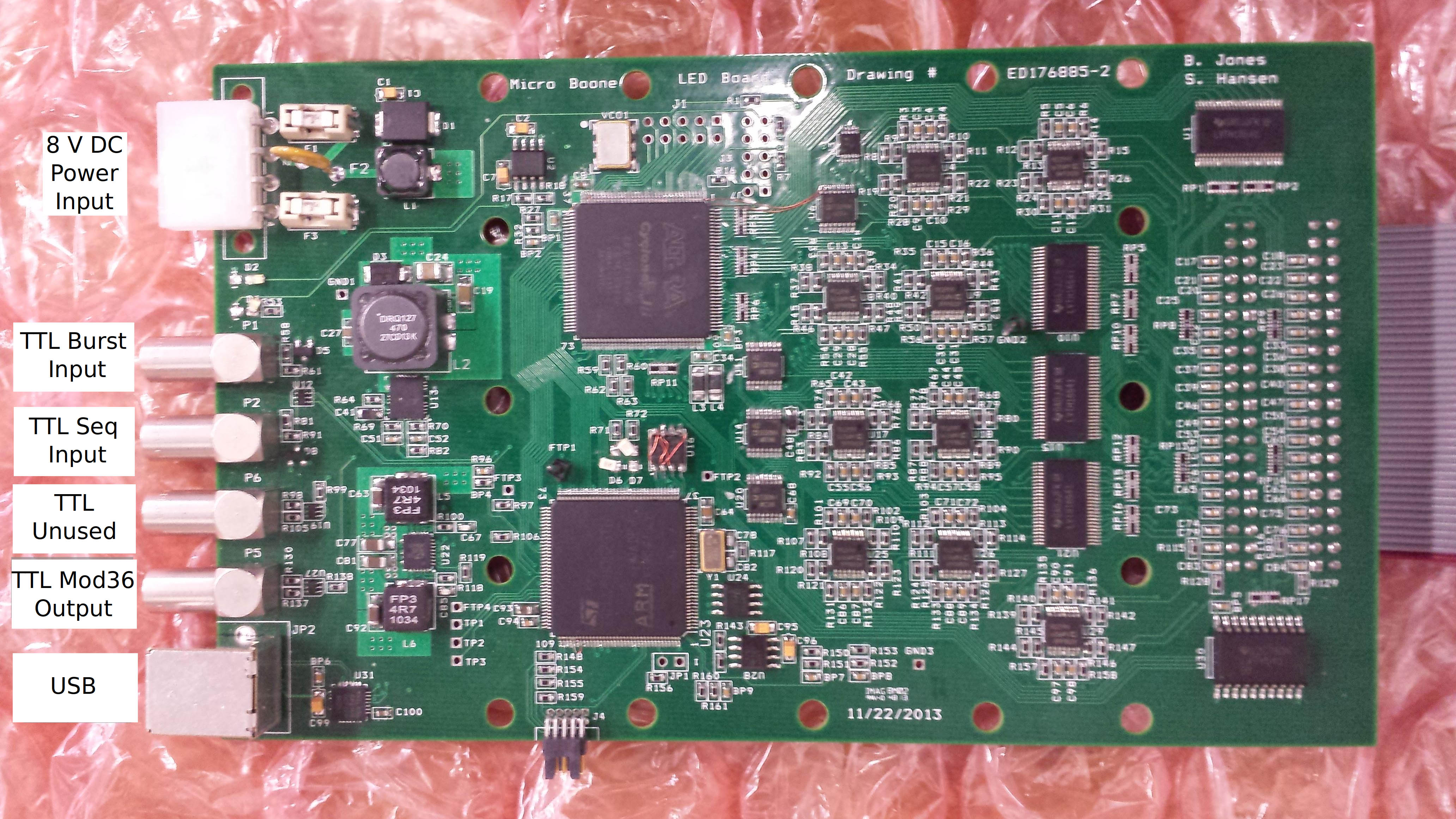}
\caption{A labeled closeup of the driver board.}
\label{fig:board_closeup_labeled}
\endminipage\hfill
\end{figure}

The 36 pulser networks of the LED board are connected to 36 outputs of the driver board (Figure \ref{fig:board_closeup_labeled}) via two keyed ribbon cables. Each of these outputs is the sum of a TTL logic driver, which provides trigger pulses, and a DAC, which provides a finely tuned DC offset voltage to control LED brightness. Both the DACs and the TTL drivers are controlled by an FPGA, which in turn is governed by a microcontroller. The microcontroller has onboard flash memory to store run configurations, as well as a USB port allowing external configuration of the flasher system. The driver board will be powered by a WIENER MPOD low voltage DC power supply.

Two LEMO connectors are provided to control triggering of the flasher system, one provides diagnostic output and one is a spare output. The first is the TTL burst mode input. Using this input, each TTL (logic high) pulse will simultaneously fire all 36 channels of the flasher system. This mode can be used to calibrate out timing differences in the PMT array, among other things. 
The second LEMO connector is the TTL sequence mode input. Repeatedly sending logic high TTL signals to this input will sequentially fire each of the 36 flasher channels. There is a range of operating frequencies (detailed below) above which glitches will cause trigger failures, and below which the sequence will reset. Sequential trigger mode can be used to perform rapid gain calibrations on each element of the PMT array and measure any potential cross-talk between light sent to different PMTs. The third LEMO connector is unused. The fourth LEMO connector is a TTL ``Mod 36'' output. This output goes high every time the 36-LED sequence is completed, providing a place-marker when operating in sequential mode with an input pulse train. The use of external triggers allows the flasher system to be operated in anti-coincidence with the Booster Neutrino Beam gate signal with the result that one can calibrate the PMT's without disturbing the collection of neutrino events. The PCB layout and schematic for this driver board can be found in the Appendix \ref{app:B} in Figures \ref{fig:LEDDRIVERPcb} and \ref{fig:LEDDriverSch} respectively.

The driver board is attached to the top of the optical stack (to the top of the protective board covering the LED board) with a set of screws. Finally, a second drilled plastic protective cover is screwed down over the back of the driver board, completing the stack. In this scheme, everything below the optical patch panel is permanently attached to the detector. Everything above, including the spacer, LED board, plastic covers, and driver board, can be replaced at any time, as they are placed outside the cryostat.

\subsection*{Measurements}
\begin{figure}[t]
\centering
\includegraphics[width=0.7\textwidth]{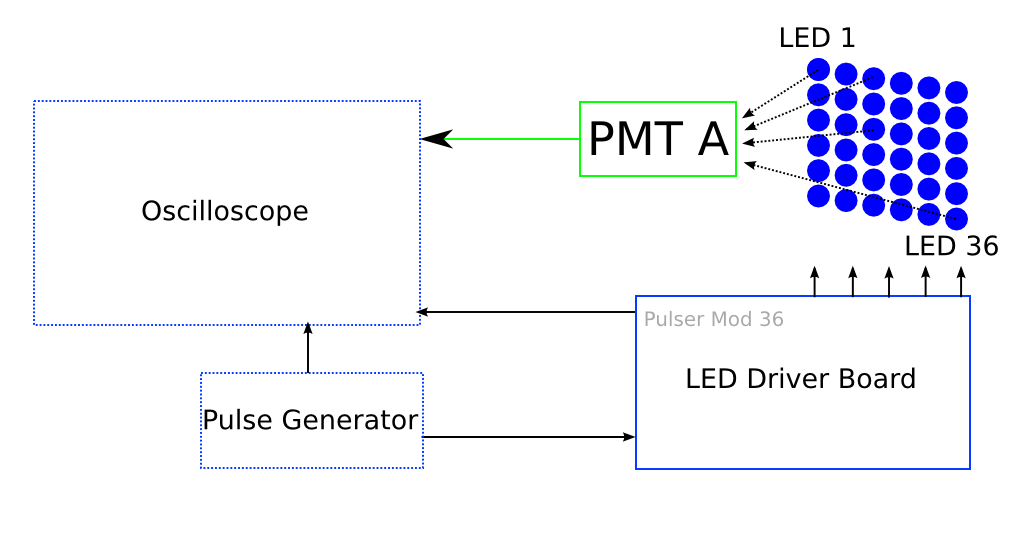}
\caption{The test setup for LED time delay measurements. }
\label{fig:image1_rotated}
\end{figure}

\begin{figure}[b]
\centering
\minipage{0.65\textwidth}
\includegraphics[width=\linewidth]{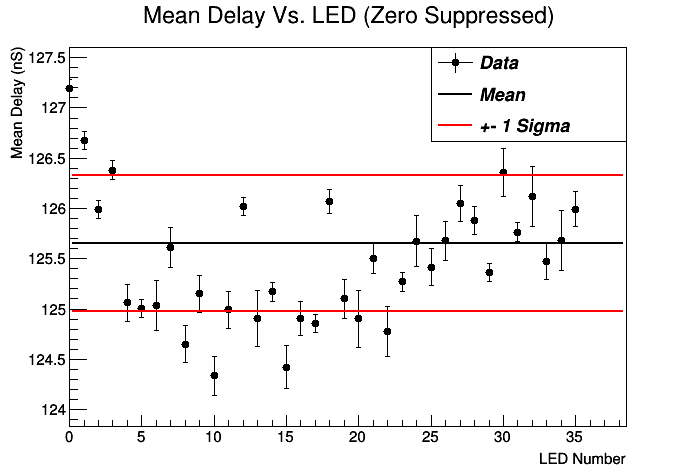}
\endminipage\hfill
\caption{Mean delay by LED number. In black is the mean delay of all LEDs weighed by associated statistical errors. The red bands show $\pm1\sigma$ standard deviations around the mean.}
\label{fig:final_mean_nofit}
\end{figure}

\begin{figure}[t]
\centering
\minipage{0.55\textwidth}
\includegraphics[width=\linewidth]{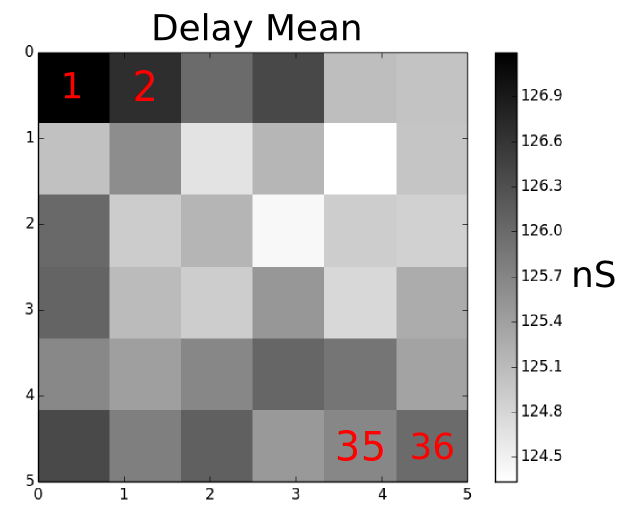}
\endminipage\hfill
\caption{Mean delay plotted by physical LED board position. There is no clear position dependence of delay times.}
\label{fig:delayfinalmap}
\end{figure}

\begin{figure}[b]
\centering
\includegraphics[width=0.7\textwidth]{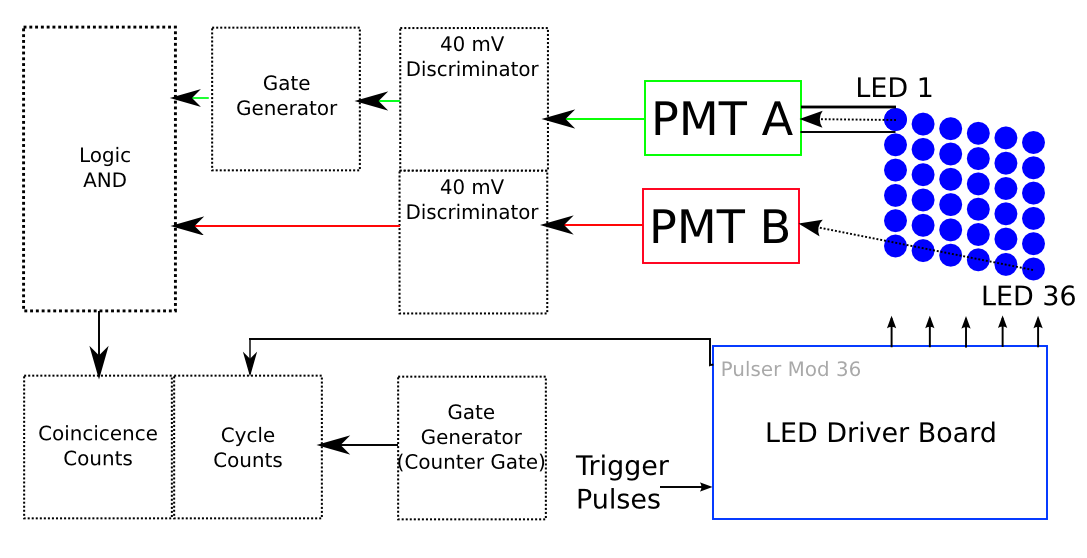}
\caption{The setup for measurement of trigger frequency dependence in the flasher system. PMT A is masked to LED1 with black vinyl. }
\label{fig:setup2}
\end{figure}

The flasher electronics were tested and characterized.  Three key metrics were measured: relative channel-to-channel trigger-signal delays, the operating frequency range, and the LED output as a function of trigger frequency. First, the delay was measured between the arrival of a TTL trigger pulse at the driver board sequential LEMO input and the rising edge of the resulting PMT signal. The apparatus used in this measurement is diagrammed in Figure \ref{fig:image1_rotated}. The LED board was positioned in front of a Phillips XP2262B PMT (``PMT A'') supplied with -1.6\,kV by a Power Designs HV-1547 and housed in a dark box. With all other LEDs off, the brightness of one LED was slowly increased until the PMT produced an output corresponding to roughly 8-16 photoelectrons (PEs). Using an Tektronix TDS5054B-NV oscilloscope, the difference in time between the rising edges of the trigger signal (supplied by an HP 8028A) and the PMT signal at 50\,\% full amplitude was measured 1000 to 2000 times per LED. For each LED, the mean and the standard deviation of the individual measurements were computed after correcting for cable delays.  The mean delay times are plotted in Figure \ref{fig:final_mean_nofit}, as a function of LED number. Using the spread in the mean delays as a measure of the systematic uncertainty gives a mean delay over all LEDs of $125.65\pm0.68$\,ns. The spread in the mean delays computed over all 36 LEDs is much larger than the uncertainty of the mean delays computed for the individual LED measurements. This could indicate real differences in the mean delay between the LEDs or an underestimate of systematic errors. Expected sources of systematic errors are the trigger cable length, the small optical path difference between different LED's in the grid and the electronics response due to small variation in the used PCB board components. We tried to isolate contributions of the systematic error, like mapping the delay time versus the position inside the 6$\times$6 matrix, as shown in Figure \ref{fig:delayfinalmap}. Since the PMT was located $\approx $5\,cm from the LED board, there could be a dependance due to the shorter travel time of LED's in the center of the matrix versus LED's on the outer positions. No clear position dependence can be observed. 

To measure the trigger frequency dependence of the LED flasher system, we modified the setup to monitor which input frequencies can be applied to successfully trigger all LED'sin the single-flash mode. We added an additional PMT to the test setup, as shown in Figure \ref{fig:setup2}, both PMT's face the LED board. One PMT, the Phillips XP2262B (``PMT A''), was supplied with -1.6\,kV from a Power Desings HV-1547, and constrained to view only the first LED '1' . A Hamamatsu R7725MOD supplied with +1.9\,kV from a Fluke 415B (``PMT B'') was able to view all the remaining 35 LEDs. As we wanted to monitor full cycles of single-flashing the 36 LED's, thus the relevant LED's to observe are the first and last of each cycle, LED '1' and '36' which were both tuned to an output corresponding to 15-30 PEs. All other LED's are irrelevant for this measurement and their pulse amplitude was thus set to zero. An internal time-out in the LED flasher board will reset the pulse sequence to LED '1' when no further trigger is applied within a sequence, such that the board will usually start with LED '1'. This was confirmed by manually pulsing the sequential trigger input; the timeout is between 10 and 15 seconds.
\begin{figure}[t]
\centering
\includegraphics[width=0.7\textwidth]{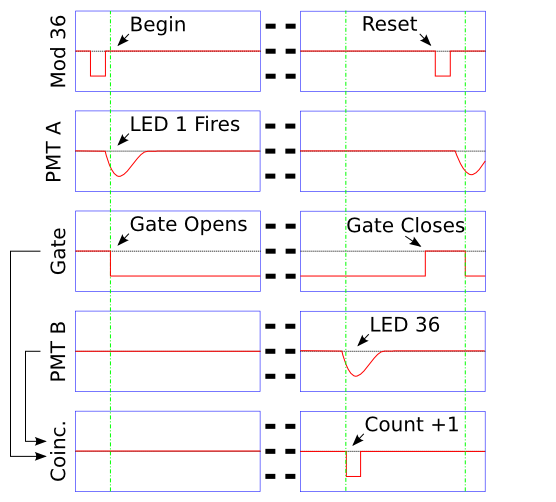}
\caption{The timing diagram for the coincidence counting. }
\label{fig:timing}
\end{figure}

\begin{figure}[t]
\centering
\includegraphics[width=0.65\textwidth]{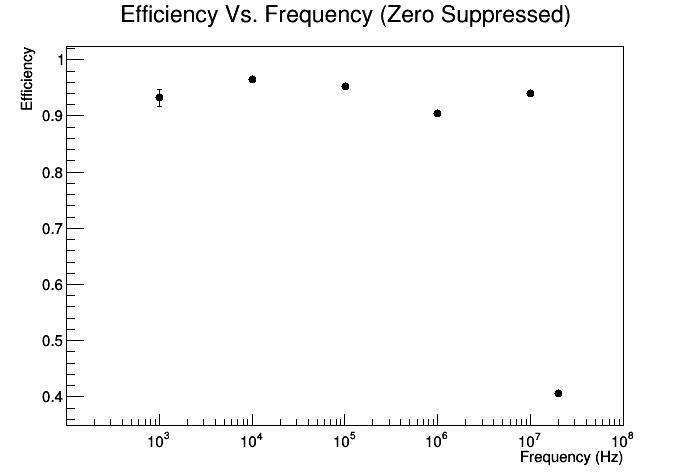}
\caption{Sequence completion efficiency as a function of trigger frequency. The efficiency is smaller than 1 due to a high discriminator threshold for the very wide pulses of the Phillips XP2262B PMT.} 
\label{fig:effplot}
\end{figure}

We used a pulse generator with a range from 1kHz to 20MHz to check for the frequency dependence. The efficiency is given by the successful pulsing of LED '36' within the timing period of the 36-LED cycle after LED '1' flashed (see Figure \ref{fig:timing}). Each pulse from the PMT viewing LED 1 opened a logic 'timing gate' with a width just below the period of one 36-LED cycle. If PMT B triggered (from LED '36') within in this time window, a ``success'' count was recorded by a scaler. Simultaneously, the number of total ``cycle counts'' was recorded by another scaler using the pulses from the LEMO ``mod-36'' output of the LED driver board (pulsed when flashing LED '1'). The ratio of ``success'' counts to cycle counts is treated as an efficiency, and is plotted against trigger frequency in Figure \ref{fig:effplot}.  The number of measured pulses was large, thus the statistical errors are negligible. The sequence is issued stable for frequencies up to 10\,MHz. The used Phillips PMT had small problems, to obtain good pulses of the LED signal one needed a high setting for the discriminator threshold. In the future, this measurement could be repeated with a different PMT. For frequencies larger than 20\,MHz, the efficiency begins to drop drastically. Investigations with an oscilloscope revealed excessive glitching: timing hiccups that cause PMT pulses to be missed and 'timing gates' to misalign.

\begin{figure}[b]
\centering
\includegraphics[width=0.65\textwidth]{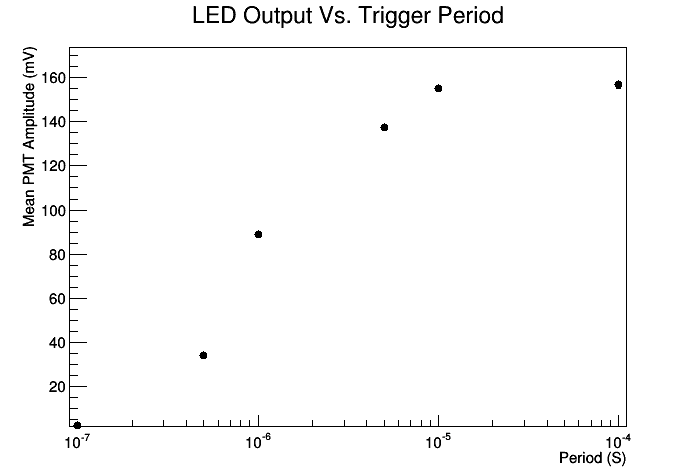}
\caption{LED amplitude as a function of pulse period.} \label{fig:rcplot}
\end{figure}

In addition to the timing glitches, high trigger frequencies cut into the charging time of the capacitor in the RC circuit that powers the LED. After the LED pulsed, its capacitance needs to be re-charged. To obtain a numerical time constant is not easy, as one can see from the line diagram in Appendix \ref{app:B} in Figures \ref{fig:LEDBOARDsch} and \ref{fig:LEDDriverSch}. For very long pulse periods, one can expect a full recharge of the capacitors, shown in Figure \ref{fig:rcplot} is the mean PMT signal amplitude as a function of trigger period for one LED at one DAC value, we investigated the behaviour of two of the LED's out of the 36. For high frequency sequences, or short trigger periods the effect of the partially charge capacitor is visible. The shape is influenced by many factors, one is that the emitted LED light is not linear to the capacitor charge, thus we refrain from comparing the obtained values to theoretical calculations of the RC circuit. The data indicate a reduced light output from the LED exists for frequencies above $O(100\,\mbox{kHz})$.  At high frequencies above 10\,MHz, significant brightness instability was observed in one of the two LEDs tested at this frequency.

\section{Conclusion} We presented the design and and construction of an LED-based fiber calibration system for large liquid argon time projection detectors. The system was designed to work within the design constraints of any LArTPC and will be used to calibrate the optical systems of the MicroBooNE experiment. We measured three key metrics: relative channel-to-channel trigger-signal delays to be within $125.65\pm0.68$\,ns, the operating frequency range was from less than 1\,Hz up to 10\,kHz and the LED output as a function of the trigger frequency shows the operational capabilities in the same interval of frequencies, all of these allowing a secure operation at the desired trigger frequency of few kHz. Detailing the materials and installation procedure, we provided technical drawings and specifications to allow the easily replication in future LArTPC detectors \cite{weblink}.

\section{Acknowledgment}
We thank Fermilab employees Sten Hansen and Terry Kiper for their vital contribution to the design and construction of the electronic circuits. We thank the workshops of LHEP Bern and Fermilab for the machining of the parts and help in design, assembly and testing of the device. All authors except Thomas Strauss are supported by National Science Foundation Grant PHY-1205175. This work was supported by the Fermilab National Accelerator Laboratory, which is operated by Fermi Research Alliance, LLC under Contract No. De-AC02-07CH11359 with the United States Department of Energy.

\newpage
\appendix
\section{Technical drawings of the LED flasher feed-through\label{app:A}}
\begin{figure}[h]
\centering
\includegraphics[width=0.8\textwidth]{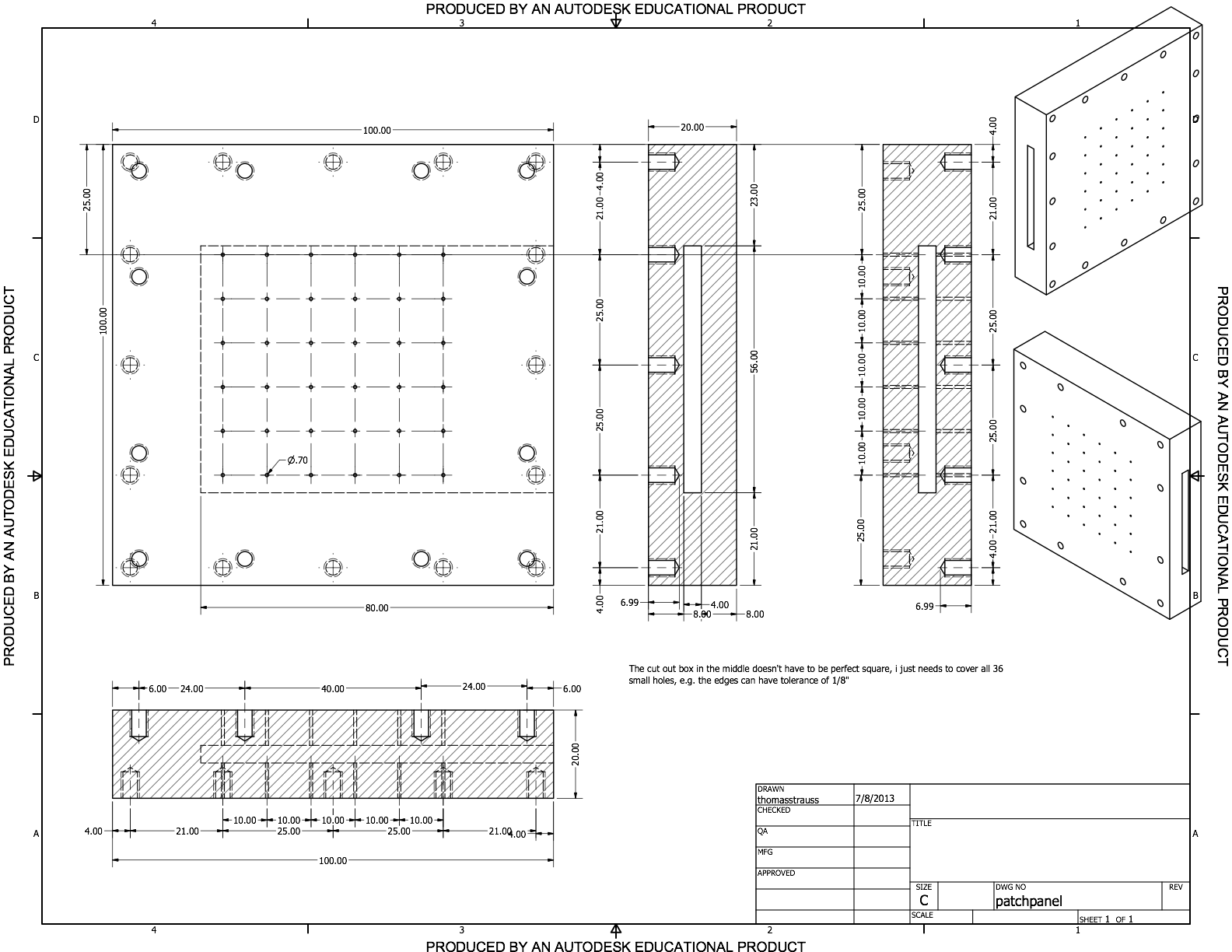}
\caption{OpticalPatchPanel. }
\label{fig:OpticalPatchPanel}
\centering
\includegraphics[width=0.8\textwidth]{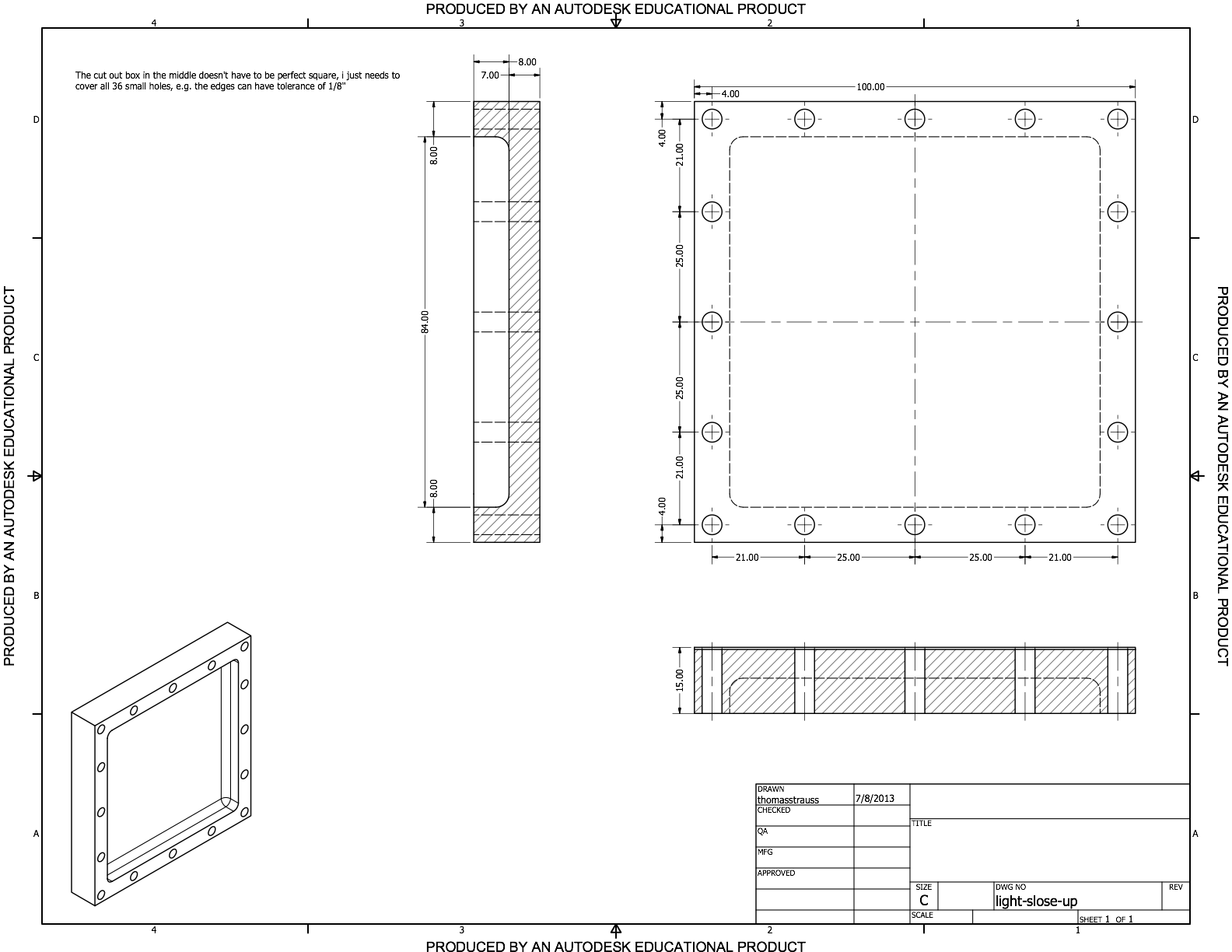}
\caption{PCBCover. }
\label{fig:PCBCover}
\end{figure}

\newpage
\begin{figure}[h]
\centering
\includegraphics[width=0.8\textwidth]{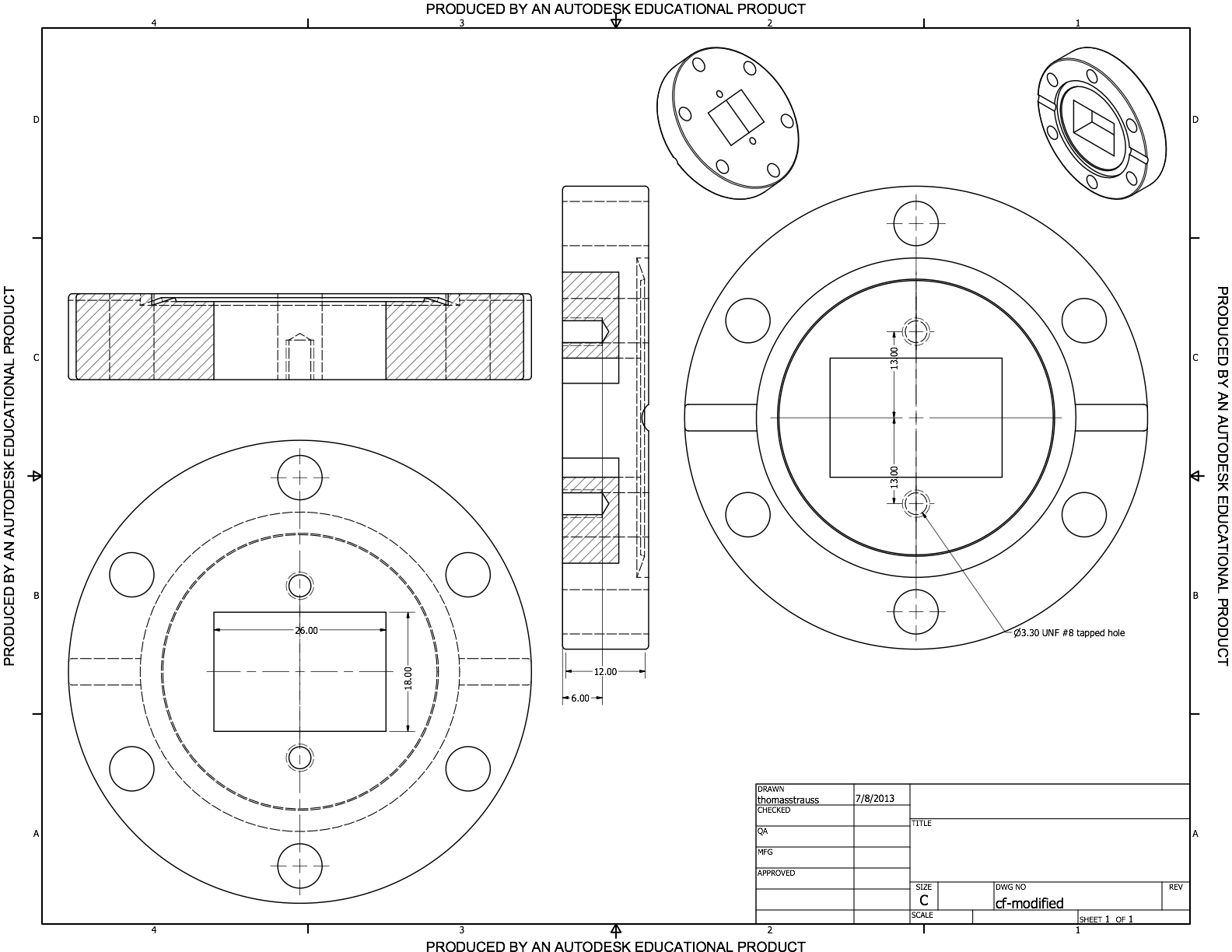}
\caption{ModifiedCFFlange. }
\label{fig:ModifiedCFFlange}
\centering
\includegraphics[width=0.8\textwidth]{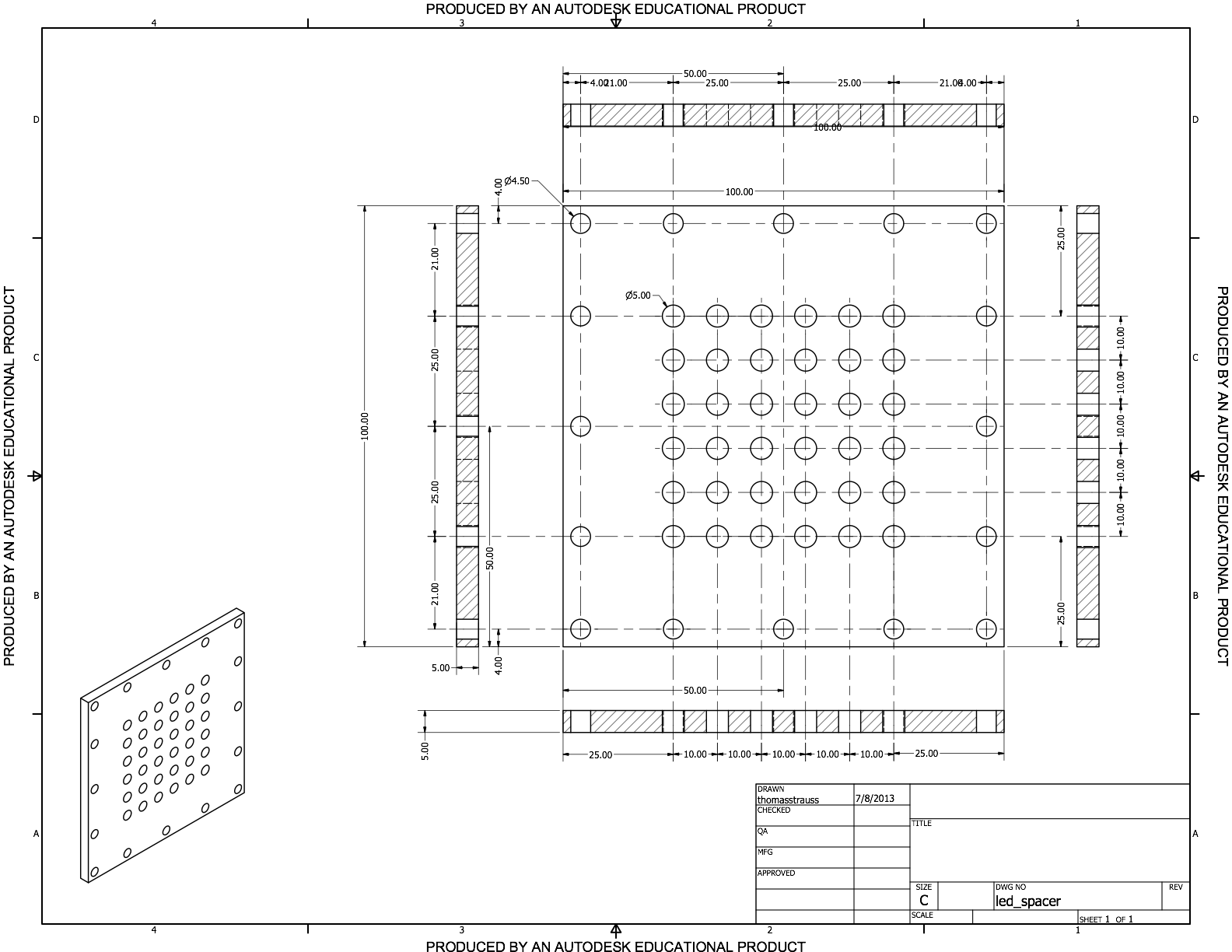}
\caption{LEDSpacer. }
\label{fig:LEDSpacer}
\end{figure}

\newpage
\begin{figure}[h]
\centering
\includegraphics[width=0.8\textwidth]{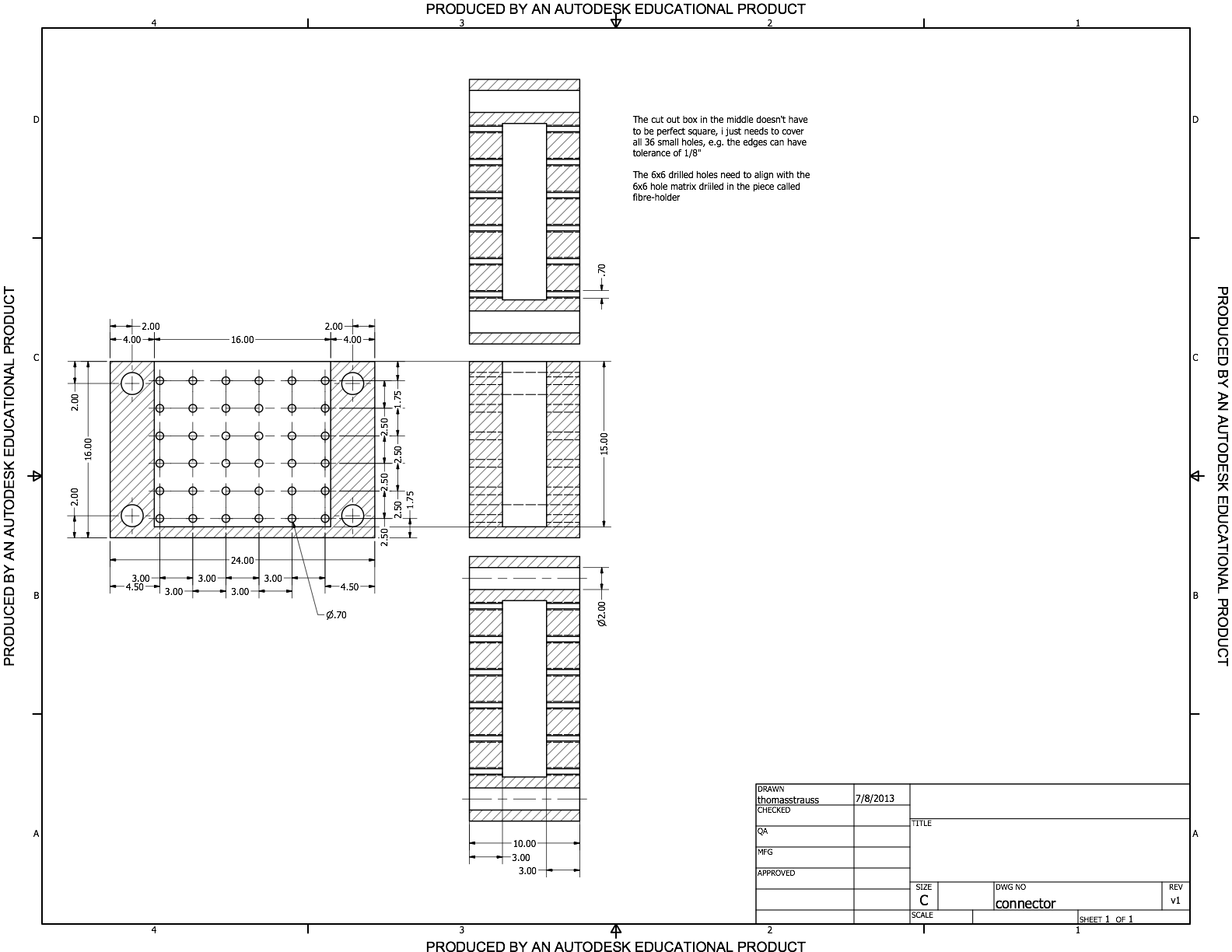}
\caption{InternalFiberConnector. }
\label{fig:InternalFiberConnector}
\centering
\includegraphics[width=0.8\textwidth]{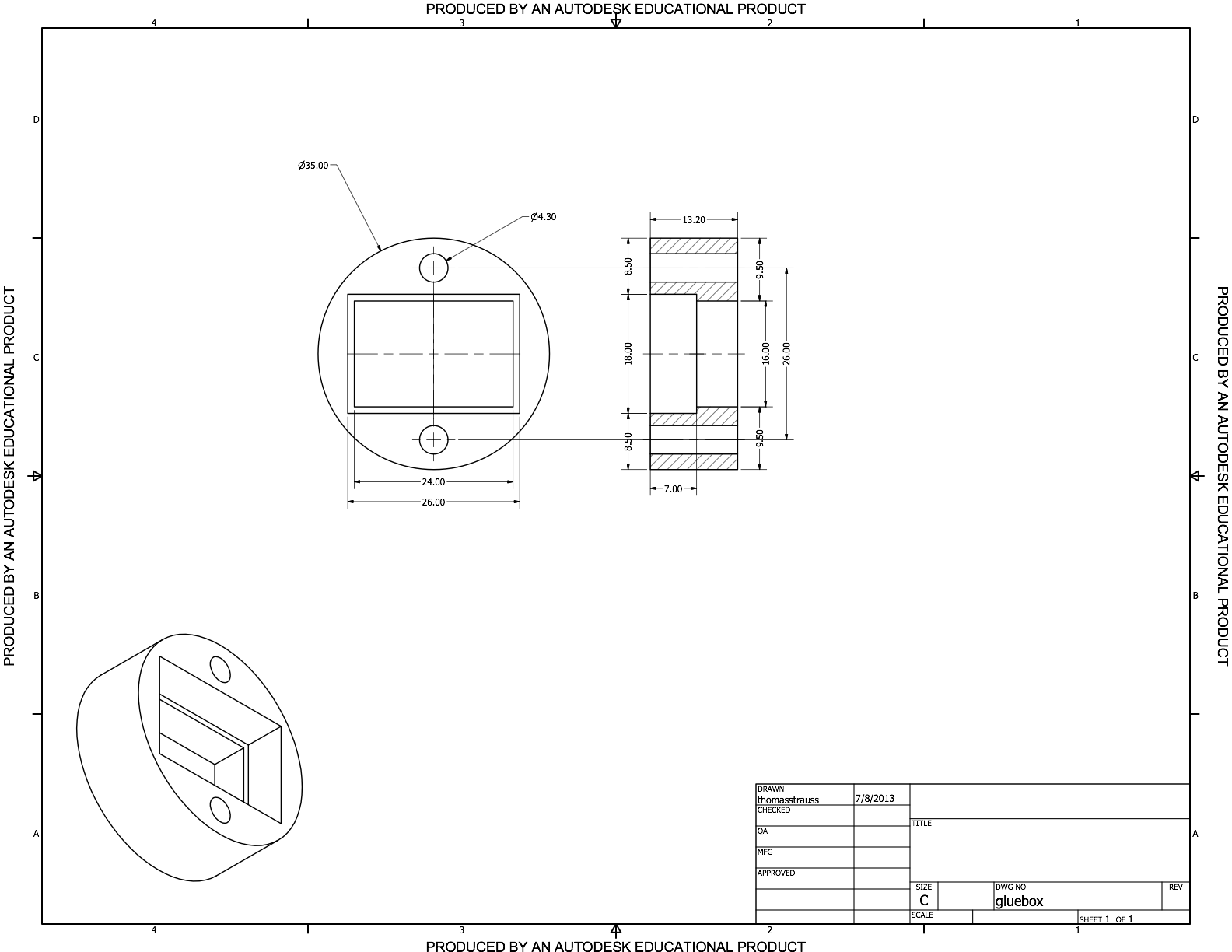}
\caption{GlueBox. }
\label{fig:GlueBox}
\end{figure}

\newpage
\begin{figure}[h]
\centering
\includegraphics[width=0.8\textwidth]{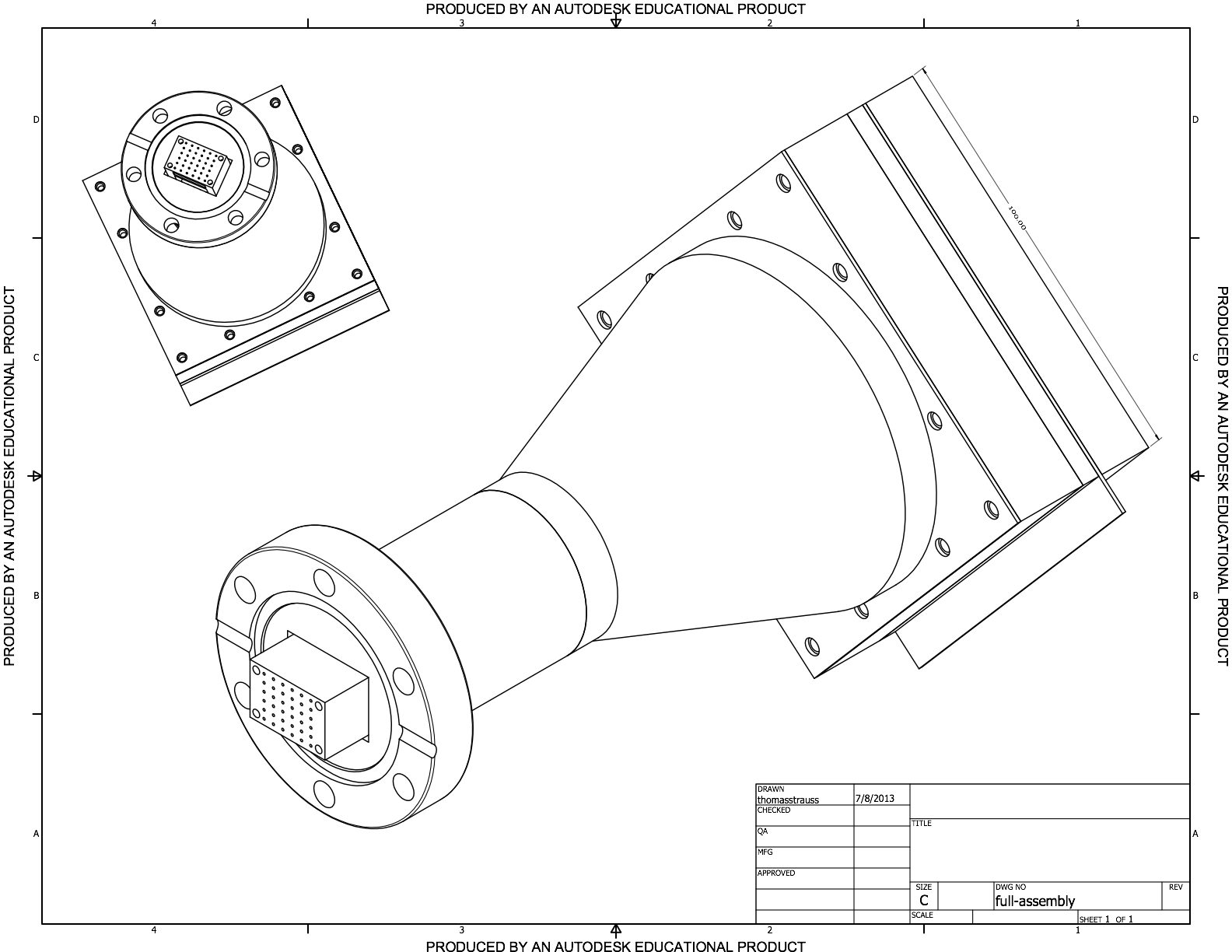}
\caption{Full feed-through. }
\label{fig:FullFeedthrough}
\centering
\includegraphics[width=0.8\textwidth]{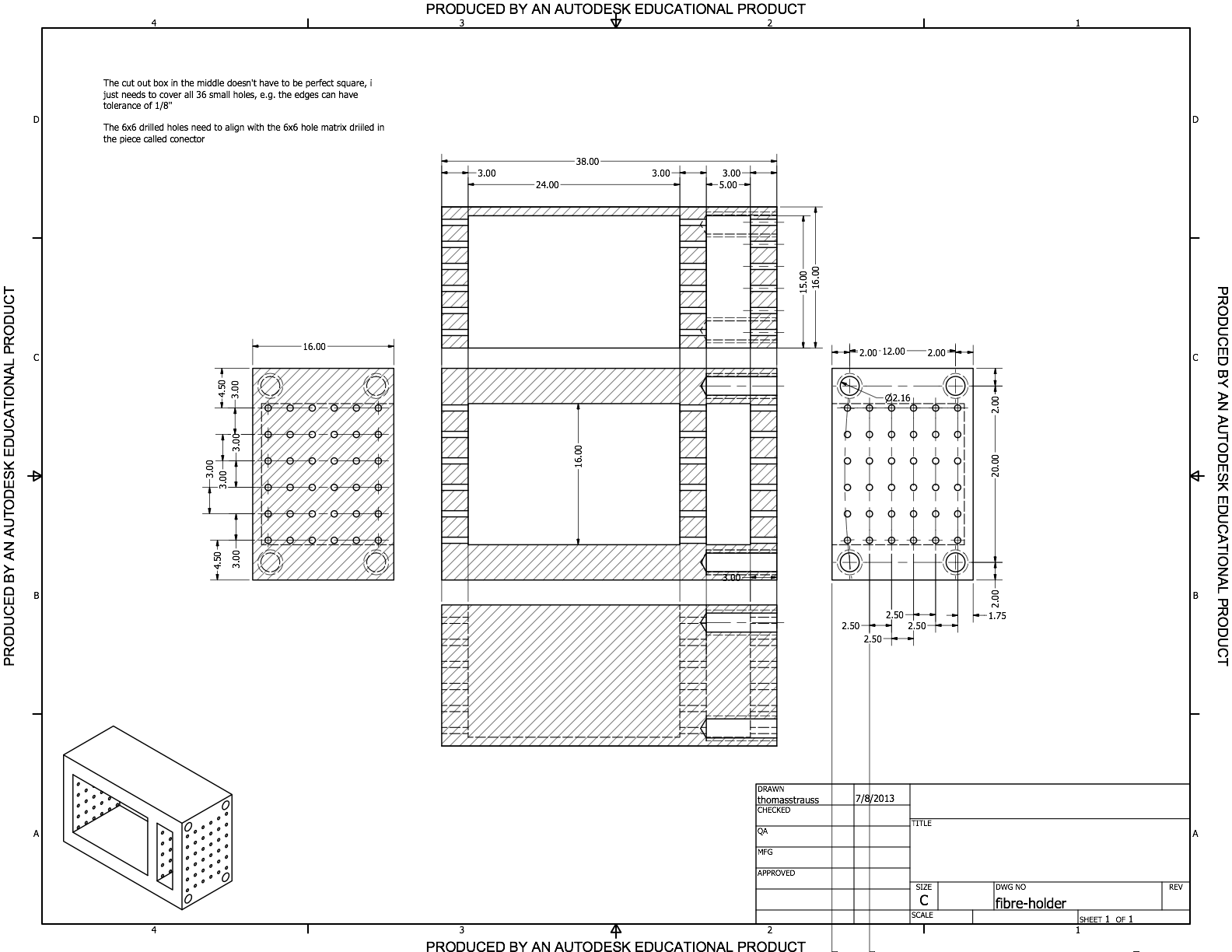}
\caption{FiberHolder. }
\label{fig:FiberHolder}
\end{figure}

\newpage
\section{Electrical drawings of the circuit boards\label{app:B}}

\begin{figure}[h]
\centering
\minipage{\textwidth}
\includegraphics[height=\linewidth,angle=90]{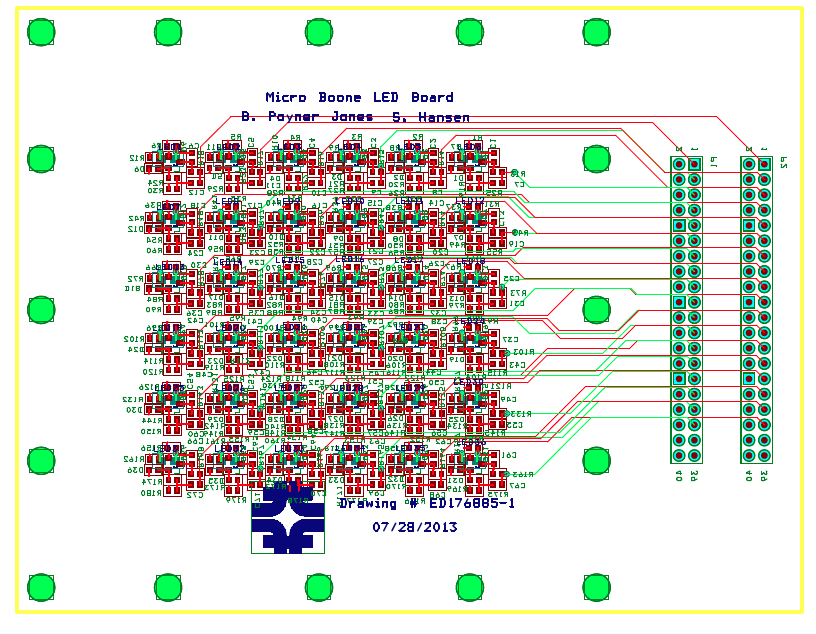}
\caption{PCB print out of the LED board. }
\label{fig:LEDBOARDpcb}
\endminipage\hfill
\end{figure}

\begin{figure}[h]
\centering
\minipage{0.95\textwidth}
\includegraphics[height=\linewidth,angle=90]{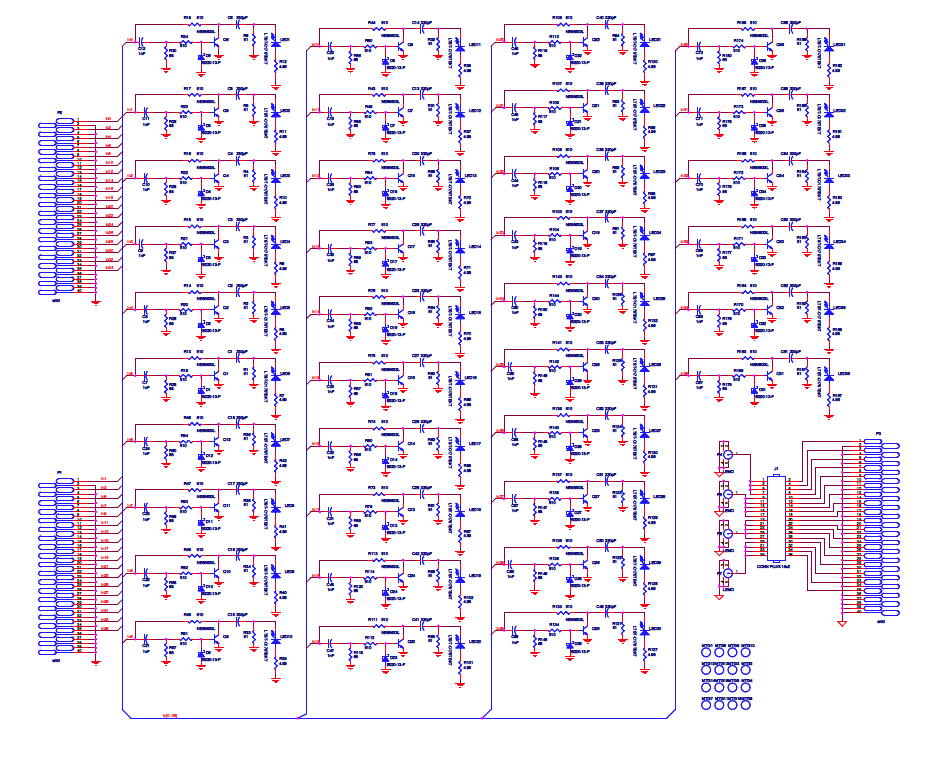}
\caption{Schematic of the LED board. }
\label{fig:LEDBOARDsch}
\endminipage\hfill
\end{figure}

\begin{figure}[h]
\centering
\minipage{\textwidth}
\includegraphics[height=0.9\linewidth,angle=90]{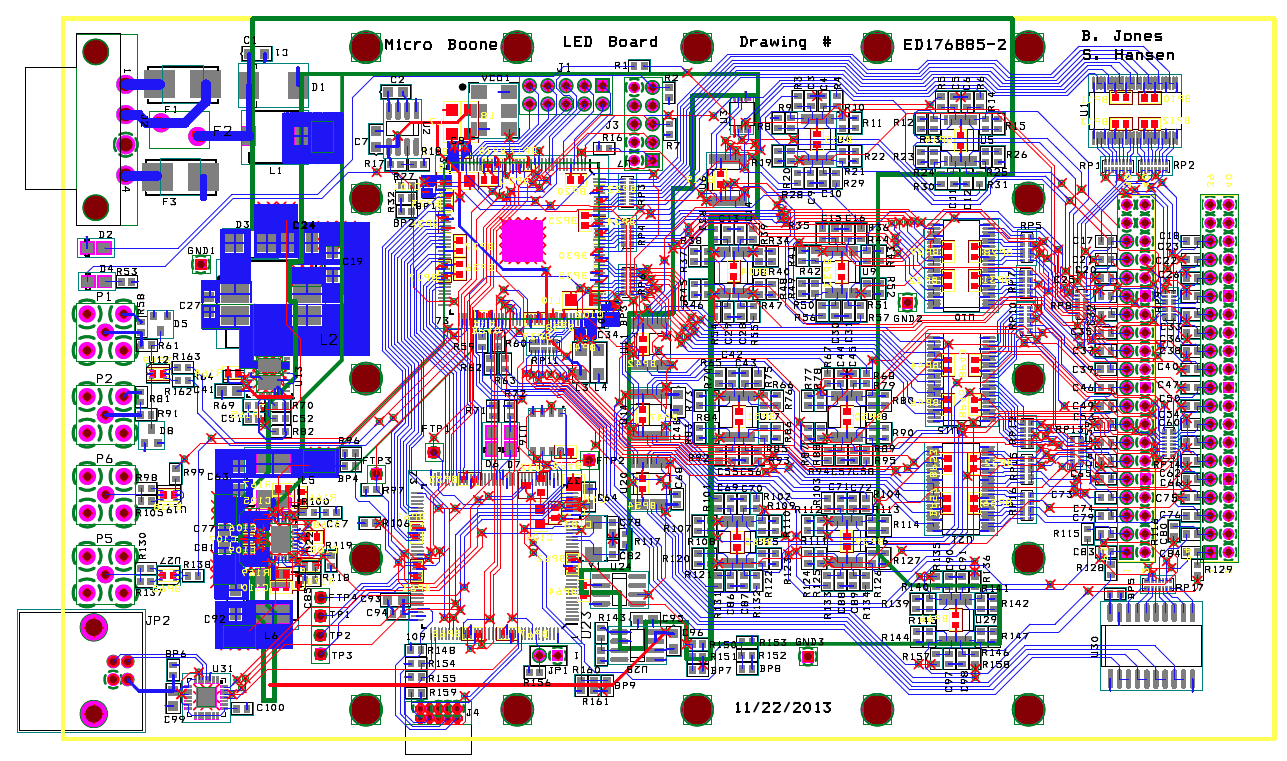}
\caption{PCB print out of the Driver board.. }
\label{fig:LEDDRIVERPcb}
\endminipage\hfill
\end{figure}

\begin{figure}[h]
\centering
\minipage{\textwidth}
\includegraphics[height=0.9\linewidth,angle=90]{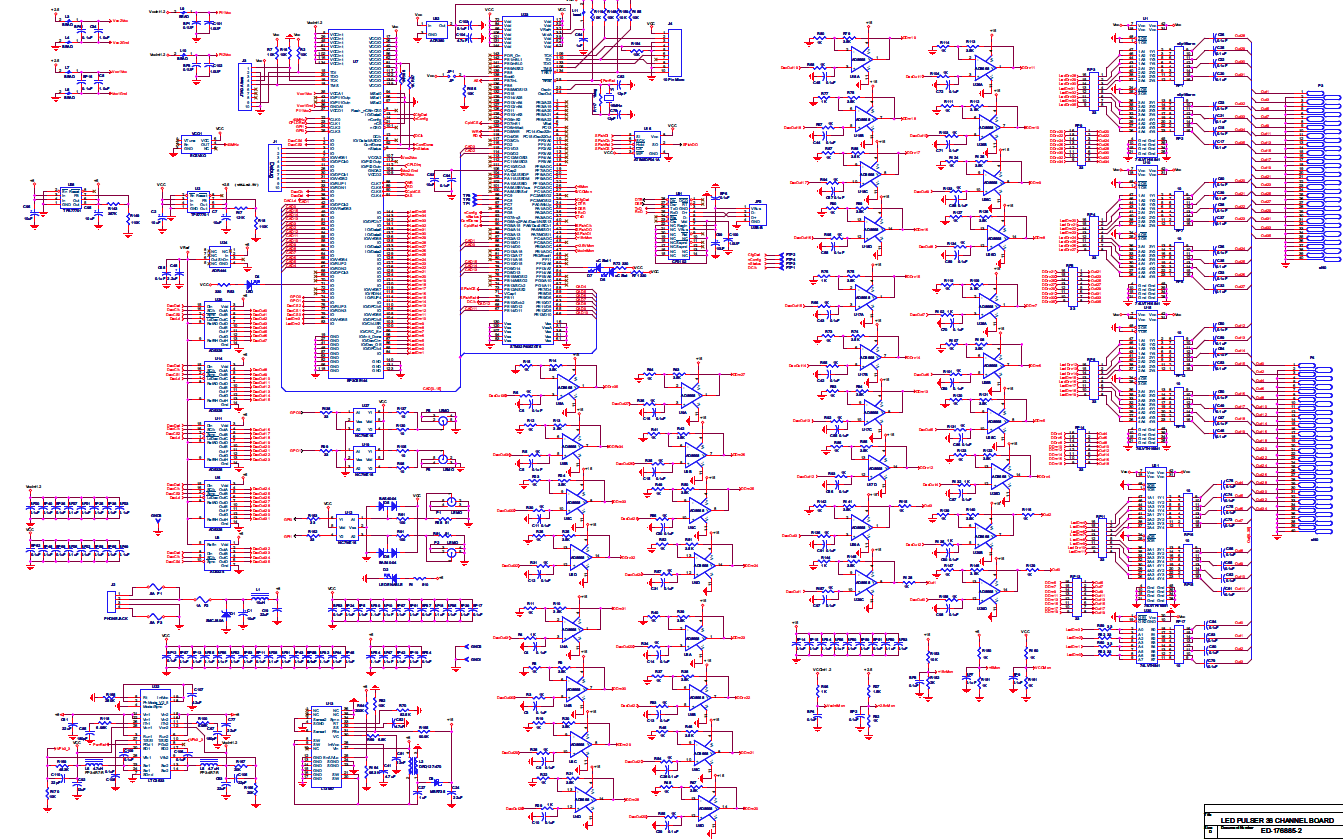}
\caption{Schematic of the Driver board. }
\label{fig:LEDDriverSch}
\endminipage\hfill
\end{figure}


\begin{thebibliography}{9}
\bibitem{Briese:2013wua} 
  T.~Briese, et al., Testing of Cryogenic Photomultiplier Tubes for the MicroBooNE Experiment, \emph{JINST} {\bf 8} T07005 (2013)

\bibitem{Baptista:2012bf} 
  B.~Baptista, et al., Benchmarking TPB-coated Light Guides for Liquid Argon TPC Light Detection Systems,  \emph{ArXiv} [physics.ins-det/1210.3793].

\bibitem{pordes} B. Rebel, et. al, Results from the Fermilab Materials Test Stand and Status of the Liquid Argon Purity Demonstrator.
 \emph{J. Phys.: Conf. Ser.} {\bf 308} (2011) 012023

\bibitem{Jones:2013bca} 
  B.~J.~P.~Jones, et al., A Measurement of the Absorption of Liquid Argon Scintillation Light by Dissolved Nitrogen at the Part-Per-Million Level,  \emph{JINST} {\bf 8} P07011 (2013)

\bibitem{cf} Conflat Flange System, \url{http://www.lesker.com/newweb/flanges/flanges_technicalnotes_conflat_1.cfm}

\bibitem{Arathane} Arathane Polyurethane Casting System, \url{http://www.farix.hu/pdf/1389712335.pdf}

\bibitem{jones} B.~J.~P.~Jones, Results from the Bo Liquid Argon Scintillation Test Stand at Fermilab, \emph{JINST} {\bf 8} C09003 (2013)

\bibitem{molex} Spec-Sheet, Molex Polymicro fibers, \url{http://www.molex.com/polymicro/opticalfibers.html}


\bibitem{Sten} Circuit board designs were created by Sten Hansen and Terry Kiper at Fermilab, Batavia, IL (USA)

\bibitem{weblink} MicroBooNE document database \url{http://microboone-docdb.fnal.gov/cgi-bin/ShowDocument?docid=3896&version=6}

\end{thebibliography}
\end{document}